\documentclass[12pt,a4paper]{article}

\usepackage[margin=25mm]{geometry}
\usepackage[numbers,sort&compress]{natbib}  
\usepackage[utf8]{inputenc}
\usepackage[T1]{fontenc}
\usepackage{hyperref}
\usepackage{url}
\usepackage{booktabs}
\usepackage{amsfonts}
\usepackage{nicefrac}
\usepackage[protrusion=true,expansion=false]{microtype}
\usepackage{xcolor}
\usepackage{amsmath}
\usepackage{graphicx}
\usepackage{float}
\usepackage{eurosym}
\usepackage{multirow}

\begin{document}

\title{The Human Utility Factor: A Computable Welfare Metric That Reframes
AI Governance as a Constrained Optimisation Problem}

\author{%
  Sivasathivel Kandasamy \\
  Independent Researcher \\
  \texttt{sivasathivel@yahoo.com}
}

\maketitle

\begin{abstract}
Existing AI governance frameworks --- including the EU AI Act, NIST
AI RMF, and thirty further instruments surveyed here --- share a
structural blind spot: none operationalises a quantitative constraint
on macro-socioeconomic stability.
A system can pass every conformity assessment while, through aggregate
labour displacement, pushing the Gini coefficient past a stability
threshold, eroding household purchasing power, and undermining the
tax base that funds social safety nets.
We formalise this gap and propose the \emph{Human Utility Factor}
(HUF), a differentiable scalar welfare metric that decomposes into
three multiplicative components --- Agency~($A$), Wellbeing~($W$),
and Economic Stability~($E$) --- each grounded in peer-reviewed
parameter estimates and expressed in terms of three observable,
actionable levers: automation depth $h_a$, redistribution intensity
$\alpha{+}\beta$, and employment coverage $\rho$.
HUF provides a closed-form interior optimum $h_a^*$ and a
redistribution threshold $(\alpha{+}\beta)^*$ below which no
automation level is welfare-positive, converting the abstract
principle that ``AI should benefit humanity'' into a triplet of
jurisdiction-specific, computable, and enforceable inequalities.
We validate HUF through a three-agent MARL simulation across three
policy regimes (BAU/U.S., Partial/Canada, Nordic) using both
analytical heuristic agents (MAE on $h_a^*$\,$\leq$\,0.20\,hrs/wk)
and independent PPO agents, which discover a Wellbeing-dominant
alternate optimum --- revealing that welfare metrics without a
separately enforced redistribution floor can be optimised into
high-automation, low-redistribution equilibria that satisfy the
metric while violating its normative intent.
The empirical evidence in Section~3 shows that the governance gap is
not hypothetical: AI-attributed layoffs in the United States grew
thirteen-fold between 2023 and 2025, GAGI has diverged from GDP
per capita in every G7 economy, and the labour share continues its
multi-decade decline.
The central reframing offered by HUF is that AI governance is not
a compliance checklist --- it is a constrained optimisation problem
with a computable objective function, three enforceable levers,
and publicly measurable parameters.
HUF is the instrument that makes this optimisation tractable: it
detects when a stability boundary is being approached and enables
course correction before the crossing becomes irreversible.
\end{abstract}

\section{Introduction}
\label{sec:intro}
Transformer based large language models (LLMs) have largely changed the landscape of technology. With its near human capabilities, one of the major impact to the society human labor displacement at unprecedented scale and unprecedented drive to automate. Many jobs have already been displaced and many more are on the way. In March 2026, artificial intelligence became the single leading cause of announced layoffs in the United States, accounting for roughly one in four of the 60,620 job cuts reported that month alone~\citep{challenger_march2026}. This is not an anomaly.  Over the full year of 2025, employers attributed approximately 55000 separations directly to AI-driven restructuring -- a nine fold increase over the prior year. While the technology sector posted its worst first-quarter job-cut total since 2001, with 52,050 announced cuts representing a 40\% year-on-year rise~\citep{challenger_march2026,fortune_cfo2026}. Meanwhile, firms executing these reductions simultaneously reported record revenues: Microsoft cut more than 15,000 roles in 2025 while posting \$70.1 billion in quarterly revenue~\citep{finalround2025}, and UPS eliminated 48,000 positions under an explicit ``automation dividend'' rationale while increasing the share of volume processed through automated facilities from 63\% to 66\%~\citep{cnbc_layoffs2025}. The pattern is economically coherent and socially destabilizing. Firms internalize the productivity gains of automation while externalizing its displacement costs onto workers, households, and public transfer systems. This is not a novel observation: \citet{acemoglu2018race} formalized it as the displacement-- reinstatement race, and \citet{acemoglu2020robots} provided causal evidence that each additional robot per 1,000 workers reduces the employment-to-population ratio by $\sim$0.2 percentage points and wages by $\sim$0.42\%. What is novel --- and alarming --- is the \emph{speed} and \emph{breadth} of the current transition. \citet{eloundou2024gpts} estimate that approximately 80\% of U.S. workers have at least 10\% of their tasks exposed to large language models, with high-wage, high-education roles now more exposed than the routine manual occupations historically targeted by automation. The IMF finds that 60\% of jobs in advanced economies face AI exposure, with roughly half at risk of displacement rather than complementarity~\citep{cazzaniga2024genai}. \citet{acemoglu2024simple} bounds the aggregate TFP gain from current AI at under 0.66\% cumulative over a decade --- a gain too small to offset the distributional losses unless explicitly redistributed. Whether the resulting productivity gains will diffuse broadly or accrue narrowly is an empirical question that current governance frameworks are not instrumented to answer.

\paragraph{The governance response is structurally inadequate.}
The international community has responded with an unprecedented proliferation of AI governance instruments: the EU AI Act~\citep{eu_ai_act_2024}, the NIST AI Risk Management Framework~\citep{nist_airmf_2023}, the UNESCO Recommendation on AI Ethics~\citep{unesco_ai_ethics_2021}, the G7 Hiroshima Process Code of Conduct~\citep{g7_hiroshima_coc_2023}, and more than two dozen national and corporate frameworks surveyed in the companion review~\citep{our_arxiv_review}. Yet, a systematic analysis of these instruments reveals a structural blind spot: \emph{no framework provides a quantitative, enforceable macro-socioeconomic stability threshold} --- frameworks that address labor or societal impact do so in aspirational or qualitative terms, without operationalized monitoring obligations or enforcement triggers tied to aggregate welfare. Human oversight provisions (e.g., EU AI Act Article~14) remain qualitative~\citep{fink_human_oversight_2025}. Risk taxonomies classify systems by deployment context, not by their aggregate labor-market impact. Socioeconomic displacement appears nowhere in enforcement mechanisms~\citep{veale2021demystifying,laux2024trustworthy}. The governance canon treats automation as a feature to be managed at the model level; it has no vocabulary for automation as a \emph{systemic macroeconomic risk}.

\paragraph{This paper's argument.}
We argue that this gap is not an oversight to be patched with better guidelines --- it is a fundamental architectural failure. Current frameworks are designed to prevent AI from doing specific harmful things (discriminating, deceiving, endangering). They are not designed to prevent AI from destabilizing the macroeconomic substrate that sustains labor-market stability and household purchasing power. Concretely: a system can pass every EU AI Act conformity assessment and every NIST RMF control, and still --- through aggregate labor displacement --- push the Gini coefficient of an economy past a stability threshold, erode the purchasing power of the median household, hollow out the middle-skill labor market, and undermine the tax base that funds the very social safety nets intended to absorb displacement shocks. This is not a hypothetical. Young workers aged 22--25 in the most AI-exposed occupations have already experienced a 13\% employment decline since ChatGPT's release in November 2022~\citep{dallasfed2026}. \citet{karabarbounis2014global} document that the labor share of income has declined in the majority of countries since the 1980s; AI threatens to accelerate this trend precisely in the decade when labor markets have the least capacity to absorb it.

\paragraph{Contributions.}
This paper makes three contributions:

\begin{enumerate}

\item \textbf{The Human Utility Factor (HUF).}
We derive a differentiable scalar welfare metric that formalises macro-socioeconomic stability constraints for AI governance (\S\ref{sec:framework}).
HUF decomposes into three multiplicative components --- Agency~($A$), Wellbeing~($W$), and Economic Stability~($E$) --- each grounded in peer-reviewed parameter estimates and expressed in terms of three observable, enforceable levers: automation depth~$h_a$, redistribution intensity~$\alpha{+}\beta$, and employment coverage~$\rho$.
The multiplicative structure enforces joint adequacy across all three components and yields a closed-form interior optimum~$h_a^*$ and a redistribution threshold~$(\alpha{+}\beta)^*$ below which no automation level is welfare-positive, converting the abstract principle ``AI should benefit humanity'' into three jurisdiction-specific, computable, and enforceable inequalities.

\item \textbf{Governance as a constrained optimisation problem.}
We implement a three-agent Stackelberg multi-agent reinforcement learning (MARL) simulation across three policy regimes calibrated to real economies (BAU/U.S., Partial/Canada, Nordic/Netherlands--Sweden), contrasting analytically-derived heuristic agents with independent PPO agents trained purely from reward feedback (\S\ref{sec:simulation}).
A 75,000-point decision-space sweep characterises the full $(h_a, \alpha{+}\beta, \rho)$ feasible region and the three-objective Pareto front.
Analytical agents validate the closed-form optimum (MAE\,$\leq$\,0.20\,hrs/wk on~$h_a^*$); neural PPO agents discover a distinct Wellbeing-dominant alternate optimum, revealing that composite welfare metrics without an independently enforced redistribution floor can be optimised into high-automation, low-redistribution equilibria that satisfy the metric while violating its normative intent.

\item \textbf{HUF as an enforcement instrument.}
HUF's three governance levers --- $h_a^*$, $(\alpha{+}\beta)^*$, and $\rho_{\min}$ --- are computable from publicly available statistics (World Bank PIP, OECD ALMP expenditure, IMF WEO), enabling jurisdiction-specific compliance thresholds (\S\ref{sec:conclusion}).
We demonstrate concretely how these translate into existing instruments: a macro-stability annex to the EU AI Act and a computable scalar replacing the NIST AI RMF's qualitative human-oversight checklist.
Unlike any instrument in Table~\ref{tab:frameworks}, HUF provides a metric that can be monitored continuously, violated detectably, and enforced prospectively before stability thresholds are breached.
\end{enumerate}

\paragraph{What this paper is not.}
This is not an argument against automation per se. Automation has historically raised living standards, extended human capability, and freed labor from dangerous and repetitive tasks~\citep{mokyr1990lever,autor2015why}. We do not claim that current AI deployment will necessarily destabilize economies --- \citet{fortune_cfo2026} correctly notes that AI-attributed layoffs remain a fraction of total job losses. Our claim is narrower and more precise: \emph{the absence of a formal stability constraint in AI governance frameworks means that society is navigating an accelerating automation transition with no instrument to detect when a stability boundary has been crossed, and no mechanism to enforce course correction before the crossing becomes irreversible.} HUF is a candidate instrument for filling that gap.

\paragraph{Structure of the paper.}
Section~\ref{sec:gaps} situates HUF in the intersection of AI governance, labor economics, and welfare-measurement literature. Section~\ref{sec:econ_impact} presents empirical evidence on the macro-socioeconomic impact of AI-driven labor displacement, including Gini trends, purchasing-power trajectories, and forward projections. Section~\ref{sec:framework} derives the HUF equation in full, with each sub-function grounded in peer-reviewed literature and operationalized in terms of measurable quantities. Section~\ref{sec:simulation} reports simulation results across governance scenarios. Section~\ref{sec:conclusion} discusses implications for framework designers, policymakers, and future work.


\section{Gaps in Current AI Governance Frameworks}
\label{sec:gaps}
The introduction asserted that existing governance instruments share a structural blind spot: no framework provides a quantitative, enforceable macro-socioeconomic stability threshold. This section substantiates that claim with specificity. We survey more than thirty governmental, intergovernmental, corporate, and civil-society frameworks along six dimensions~\citep{our_arxiv_review} and show that none provides an operationalized, quantitative constraint on aggregate welfare outcomes (Table~\ref{tab:frameworks}). We then identify three structural gaps --- the \emph{socioeconomic protection gap}, the \emph{formal-metric gap}, and the \emph{feedback-loop gap} --- and argue that these gaps are not accidental omissions but an architectural consequence of frameworks being designed around model-level risk rather than system-level stability.

\subsection{A Taxonomy of Existing Frameworks}
\label{sec:taxonomy}
Governance of AI has stratified into four layers: binding law, voluntary standards, corporate self-regulation, and civil-society accountability instruments. The most significant binding instrument is the EU AI Act~\citep{eu_ai_act_2024}, which establishes a four-tier risk pyramid (unacceptable / high / limited / minimal) with penalties up to \euro~35M or 7\% of global annual turnover, and designates general-purpose AI models exceeding $10^{25}$ FLOPs as presumptively systemic. The Council of Europe Framework Convention~\citep{coe_cets225_2024} extends binding obligations to a treaty form.
At the voluntary standard level, the NIST AI Risk Management Framework~\citep{nist_airmf_2023} organizes obligations into four functions (\textsc{Govern}, \textsc{Map}, \textsc{Measure}, \textsc{Manage}) and remains the dominant reference architecture for U.S. industry compliance. The UNESCO Recommendation~\citep{unesco_ai_ethics_2021}, OECD AI Principles~\citep{oecd_ai_principles_2019}, and G7 Hiroshima Process Code of Conduct~\citep{g7_hiroshima_coc_2023} operate as soft-law convergence instruments without enforcement authority. Corporate self-governance has produced capability-threshold policies: Anthropic's Responsible Scaling Policy~\citep{anthropic_rsp_2024}, OpenAI's Preparedness Framework~\citep{openai_preparedness_2025}, DeepMind's Frontier Safety Framework~\citep{deepmind_fsf_2024}, and Meta's Frontier AI Framework~\citep{meta_frontier_2025} all define ordinal risk tiers keyed to dangerous-capability evaluations, with internal review boards as the sole enforcement mechanism.

Table~\ref{tab:frameworks} maps thirty-two of these instruments across six dimensions: binding status, risk taxonomy, compute thresholds, human-oversight specification depth, enforcement mechanism, and socioeconomic displacement coverage. The pattern is consistent: no framework provides a quantitative, enforceable macro-socioeconomic stability threshold. Frameworks that address labor at all do so either by designating employment decisions as a high-risk \emph{use case} (EU AI Act Annex~III) or by noting in aspirational language that AI ``should benefit humanity.'' Neither constitutes an operationalized stability constraint with monitoring obligations or enforcement triggers tied to aggregate welfare outcomes.


\begin{table}[h]
\caption{%
  Comparative taxonomy of thirty-two major AI governance frameworks across five dimensions. \textbf{Binding}: direct legal enforceability (Yes = statute or treaty; Fed.\ = binding on federal agencies only; Self = voluntary corporate commitment; Pend.\ = pending enactment). \textbf{Risk}: explicit ordinal risk classification or equivalent (--- = none). \textbf{FLOP}: compute-based deployment threshold (--- = absent). \textbf{HO}: depth of human-oversight specification (H = high, M = medium, L = low). \textbf{Enforcement}: primary accountability mechanism. \emph{A sixth dimension --- substantive coverage of macro-socioeconomic displacement --- was evaluated for every instrument in this table. The entry is uniformly absent across all thirty-two frameworks; no instrument defines a quantitative threshold, monitoring obligation, or enforcement mechanism tied to aggregate labor-market stability. This constitutes the socioeconomic protection gap formalized in
  \S\ref{sec:gap1}.}%
}
\label{tab:frameworks}
\centering
\scriptsize
\setlength{\tabcolsep}{3pt}
\begin{tabular}{@{}p{3.6cm}lllp{2.8cm}@{}}
\toprule
\textbf{Framework} & \textbf{Binding} & \textbf{Risk} & \textbf{HO} & \textbf{Enforcement} \\
\midrule
\multicolumn{5}{@{}l}{\textit{Governmental and intergovernmental}} \\[2pt]
EU AI Act \citep{eu_ai_act_2024} & Yes & 4-tier & H & Fines up to 7\% global revenue \\
CoE CETS~225 \citep{coe_cets225_2024} & Treaty & Graded & M & National implementing law \\
US EO~14110 \citep{biden_eo14110_2023} & Fed. & --- & M & Rescinded Jan 2025 \\
US AI Bill of Rights \citep{ostp_aibor_2022} & No & --- & M & None \\
NIST AI RMF \citep{nist_airmf_2023} & No & --- & M & Voluntary \\
NIST GenAI Profile \citep{nist_genai_2024} & No & 12 risks & M & Voluntary \\
UK White Paper \citep{uk_ai_whitepaper_2023} & No & --- & L & Sectoral regulators \\
China GenAI Measures \citep{china_genai_interim_2023} & Yes & Graded & M & Filing \& fines \\
UNESCO Recommendation \citep{unesco_ai_ethics_2021} & No & --- & M & Member-state reports \\
OECD AI Principles \citep{oecd_ai_principles_2019} & No & --- & L & Policy observatory \\
G7 Hiroshima CoC \citep{g7_hiroshima_coc_2023} & No & --- & L & OECD reporting \\
Singapore MGF \citep{singapore_maig_2020} & No & $p{\times}s$ & H & Voluntary \\
Canada Dir.\ ADM \citep{canada_directive_adm_2019} & Fed. & 4-level & H & AIA score \\
Colorado AI Act \citep{colorado2024sb205} & Yes & High-risk & M & \$20K per violation \\
Brazil PL~2338 \citep{brazil_pl2338_2023} & Pend. & 3-tier & H & BRL~50M / 2\% revenue \\
ISO/IEC 42001 \citep{iso_iec_42001_2023} & No & Controls & M & Third-party certification \\
IEEE 7010 \citep{ieee70102020} & No & WIA & M & Voluntary \\[4pt]
\multicolumn{5}{@{}l}{\textit{Corporate safety frameworks}} \\[2pt]
Anthropic RSP \citep{anthropic_rsp_2024} & Self & ASL 1--5 & M & Internal; LTBT \\
OpenAI Preparedness \citep{openai_preparedness_2025} & Self & L/M/H/C & M & Internal safety advisory \\
DeepMind FSF \citep{deepmind_fsf_2024} & Self & CCL & M & Internal safety cases \\
Meta Frontier AI \citep{meta_frontier_2025} & Self & H/C & L & Internal review \\
Microsoft RAI Std.\ \citep{microsoft_rai_2022} & Self & Sensitive & M & Office of Resp.\ AI \\
Google SAIF \citep{google_saif_2023} & Self & 6-layer & L & Industry; internal \\
MLCommons AILuminate \citep{ghosh2025ailuminate} & No & 5 grades & M & Open benchmark \\[4pt]
\multicolumn{5}{@{}l}{\textit{Training and evaluation methodologies}} \\[2pt]
Constitutional AI \citep{bai2022constitutional} & Self & --- & L & AI feedback loop \\[4pt]
\multicolumn{5}{@{}l}{\textit{Civil society and academic instruments}} \\[2pt]
Asilomar Principles \citep{fli_asilomar_2017} & No & --- & L & None \\
Montr\'{e}al Declaration \citep{montreal_declaration_2018} & No & --- & L & None \\
AI Now Reports \citep{ainow2023} & No & --- & L & Advocacy \\
FLI Pause Letter \citep{fli_pause_2023} & No & --- & L & None \\
Partnership on AI \citep{pai_synthetic_2023} & No & --- & L & Signatory commitment \\
Bletchley Declaration \citep{bletchley_declaration_2023} & No & --- & L & None \\
Paris AI Statement \citep{paris2025statement} & No & --- & L & None \\
\bottomrule
\end{tabular}
\end{table}

\subsection{Three Structural Gaps, and Why They Are Architectural}
\label{sec:gap1}

Three gaps recur across all thirty-two instruments; a full per-framework
analysis is provided in the companion review~\citep{our_arxiv_review}.

\paragraph{The socioeconomic protection gap.} No framework defines a
quantitative threshold beyond which aggregate labor displacement
constitutes a governance violation, nor mandates monitoring of macro-level
indicators such as the Gini coefficient, real-wage trajectories, or
labour-force participation. This is a foundational design choice: every
instrument was built to govern \emph{a system} --- a model, a product, a
deployment --- rather than \emph{an economic transition}. The EU AI Act
designates AI-assisted hiring decisions as high-risk (Annex~III,
\citealp{eu_ai_act_2024}), but this governs the fairness of individual
algorithms, not the aggregate effect of automation on employment levels
\citep{veale2021demystifying}: a firm can lawfully comply with every
requirement while eliminating thousands of roles, and the framework has no
vocabulary for that outcome~\citep{laux2024trustworthy}.
\citet{acemoglu2019automation} and \citet{korinek2019ai} formalize this as
a structural market failure --- a productivity transfer from labour to
capital that does not self-correct without redistribution mechanisms
activated \emph{before} displacement crosses a critical threshold --- yet
no governance instrument translates this finding into a binding or even
advisory stability threshold.

\paragraph{The formal-metric gap.} Every framework in
Table~\ref{tab:frameworks} invokes ``human oversight'' or
``human-centricity'' as a governing principle, but none provides a scalar,
computable quantity that would let an auditor determine whether the
principle is satisfied. The EU AI Act's Article~14 enumerates five
qualitative oversight capabilities that \citet{fink_human_oversight_2025}
shows cannot be evaluated without secondary technical standards that do
not yet exist; the NIST AI RMF defines risk compositionally but prescribes
no formula for either operand in the labor-displacement context
\citep{nist_airmf_2023}. This has a direct consequence: enforcement is
impossible in principle, since a regulator cannot determine whether a
qualitative principle is violated without a metric to evaluate against. The
fairness literature has independently built exactly this kind of
mathematical infrastructure --- additively decomposable scalar inequality
measures \citep{speicher2018unified,shorrocks1980class} and formal
incompatibility results \citep{hardt2016equality,kleinberg2017inherent,%
chouldechova2017fair}, operationalized in open-source toolkits such as IBM
AI Fairness 360~\citep{bellamy2019aif360} --- yet none of it has been
incorporated into a binding governance instrument.

\paragraph{The feedback-loop gap.} All instruments evaluate AI systems in
isolation or at the point of deployment; none models the dynamic loop
connecting automation adoption to macro-level outcomes over time.
Automation displaces workers and depresses wages, which depresses
aggregate demand~\citep{keynes1936general}, which reduces firm revenue and
investment capacity, which weakens tax revenue and the public capacity to
fund retraining --- a self-reinforcing cycle that \citet{acemoglu2018race}
term an ``automation trap'' and that \citet{dosi2010schumpeter} formalize
in an agent-based macro model. \citet{bommasani2022picking} show that
shared-foundation-model adoption correlates these decisions across firms,
amplifying displacement shocks beyond what any individual-system
assessment would predict --- a form of algorithmic monoculture
\citep{creel2022algorithmic,kleinberg2021monoculture} with no counterpart
in existing regulatory categories.

\paragraph{Why the gaps are architectural.} All three share a common root:
every framework surveyed was designed to answer \emph{does this system
behave safely in its intended use context} --- a model-level question that
presupposes aggregate social outcomes can be inferred from individual
compliance. Displacement harm violates that presupposition on three
counts. It is \emph{emergent}, arising from the cumulative interaction of
many individually compliant deployments sharing one labour market. It is
\emph{threshold-nonlinear}: macro-stability indicators are stable across a
wide range of automation levels and then shift discontinuously once a
critical mass is crossed~\citep{korinek2019ai,acemoglu2019automation}. And
it is \emph{irreversible} on policy-relevant timescales --- tacit skills
erode exponentially once workers leave automation-exposed roles
\citep{bainbridge1983ironies,degrip2002economics}, and displaced
middle-skill workers do not recover pre-displacement earnings even a
decade later~\citep{autor2015why}. A framework built to evaluate
individual-system compliance cannot detect, prevent, or remediate this
class of harm by construction. The missing component is a system-level
stability constraint: a computable quantity that maps the aggregate state
of automation deployment to a scalar welfare index, monitorable
continuously and enforceable prospectively, before a stability threshold
is breached. This is precisely what the Human Utility Factor, derived in
Section~\ref{sec:framework}, is designed to provide.


\section{Socioeconomic Impact of Governance Gaps}
\label{sec:econ_impact}

The three gaps identified in Section~\ref{sec:gaps} are not merely
analytical deficiencies --- they have measurable, accelerating costs that
are already visible in the data. We document them with the
\emph{Gini-Adjusted GDP per Capita Index} (GAGI), an inequality- and
inflation-corrected rescaling of GDP per capita. A companion empirical
paper~\citep{gagi2026} develops GAGI in full and applies it to the G7
economies across five empirical dimensions --- per-country welfare
trajectories, AI-attributed labour displacement, the
productivity--wage divergence, inequality regimes, and forward
projections --- with five figures and a complete sourcing apparatus; we
summarise only the headline findings needed to motivate the framework
that follows.

\subsection{The GAGI Metric and Its Headline Pattern}
\label{sec:s3_country}
For country~$k$ and year~$t$:
\begin{equation}
  \mathrm{GAGI}_{k,t}
  \;=\;
  \frac{\bigl(1 - G_{k,t}\bigr)\;\times\;\mathrm{GDP\text{-}pc}_{k,t}}
       {\pi_{k,t}}
  \;\bigg/\;
  \frac{\bigl(1 - G_{k,2010}\bigr)\;\times\;\mathrm{GDP\text{-}pc}_{k,2010}}
       {\pi_{k,2010}},
  \label{eq:gagi}
\end{equation}
where $G_{k,t}\in[0,1]$ is the Gini coefficient \citep{worldbank2025pip}, $\mathrm{GDP\text{-}pc}_{k,t}$ is GDP per capita in constant 2015 USD \citep{worldbank2025wdi}, and $\pi_{k,t}$ is the CPI index (2010\,=\,100) \citep{worldbank2025wdi}, all normalised to a 2010 baseline so that values above (below) 1.0 indicate inequality-adjusted, inflation-corrected prosperity above (below) its 2010 level. Dividing already-deflated GDP per capita by the CPI is a deliberate modelling choice --- the GDP deflator and CPI price different baskets, and the wedge between them is empirically meaningful --- discussed and defended in detail in the companion paper \citep{gagi2026}. GAGI is a simplified empirical analogue of the Economic Stability sub-function $E(\cdot)$ derived in Section~\ref{sec:framework}.

\begin{figure*}[t]
  \centering
  \includegraphics[width=\textwidth]{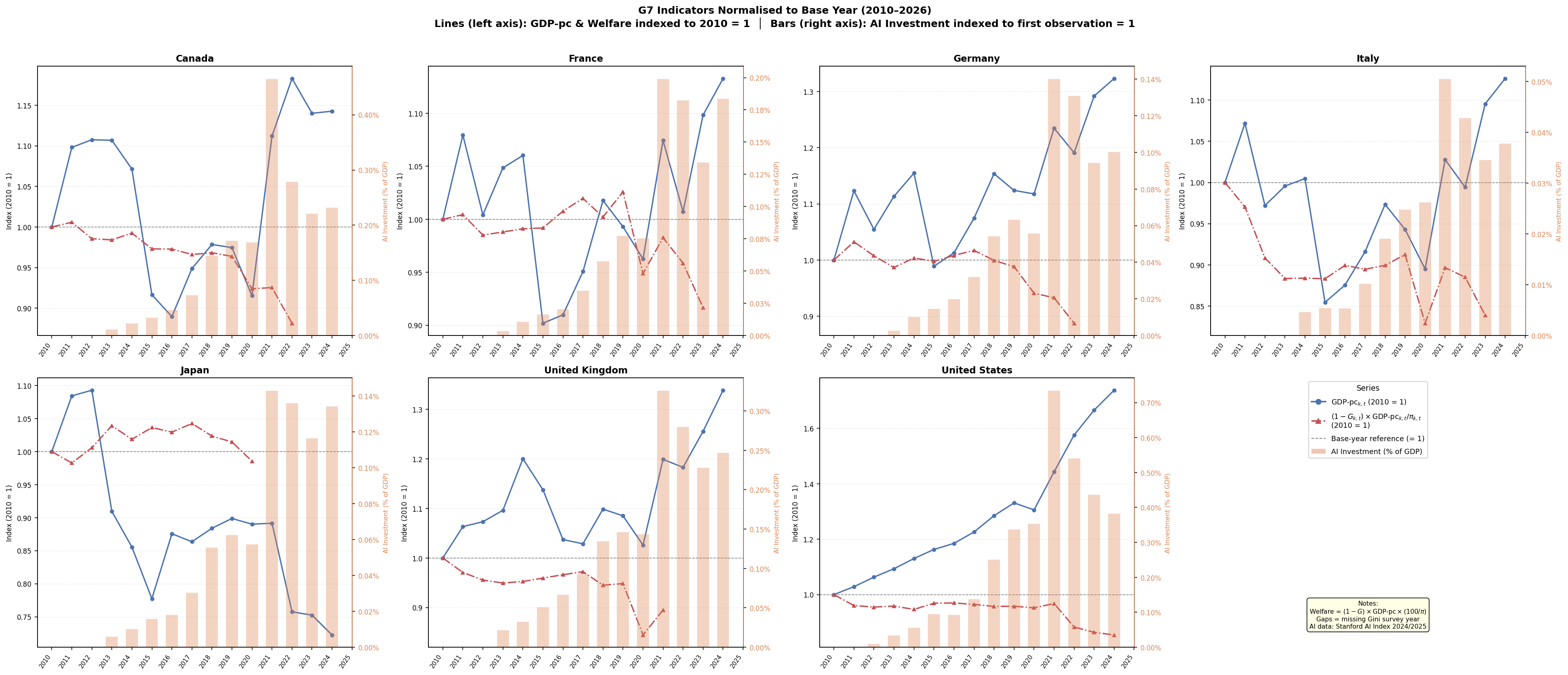}
  \caption{%
    \textbf{Gini-Adjusted GDP per Capita Index (GAGI), GDP per
    Capita Index, and Private AI Investment, G7 Economies,
    2010--2026 (2025--2026 preliminary).}
    \textbf{Observation.}
    The GDP per capita index (blue line, left axis) consistently exceeds
    the GAGI index (red line, left axis) in every G7 country throughout
    the period, with the gap widening post-2022, coincident with
    accelerating AI investment (bars, right axis, country-specific scale).
    GAGI\,=\,$(1{-}G)\times\mathrm{GDP}_{pc}\,/\,\pi$,
    normalised to 2010\,=\,1.0 (Eq.~\ref{eq:gagi}).
    \textbf{Interpretation.}
    The persistent divergence quantifies the distributional cost that raw GDP
    growth conceals: productivity gains are accruing faster than they reach
    households once rising inequality and inflation are accounted for.
    The United States shows the largest and most sustained gap; Japan and Italy
    exhibit near-flat GAGI despite low AI investment, exposing structural
    stagnation as a distinct welfare failure mode.
    \textbf{Implication.}
    Governance instruments calibrated to GDP rather than GAGI-type metrics
    will systematically underestimate the welfare cost of automation-driven
    distributional harm --- the formal-metric gap (\S\ref{sec:gap1}) rendered
    empirically visible.
    Sources: \citet{worldbank2025pip,worldbank2025wdi,%
    imf2025weo,imf2026weo,aiindex2024,aiindex2025}.%
  }
  \label{fig:country_subplots}
\end{figure*}

Figure~\ref{fig:country_subplots} presents GAGI (solid line), the GDP per capita index (dashed line), and private AI investment as a share of GDP (bars) for each G7 economy, 2010--2026 (2025--2026 values are preliminary/partial-year IMF estimates~\citep{imf2025weo,imf2026weo}).

In every G7 economy the GDP-per-capita index lies above the GAGI index throughout 2010--2026, and the wedge has widened sharply post-2022 (Figure~\ref{fig:country_subplots}): raw economic growth consistently overstates welfare-adjusted prosperity once inequality and inflation are accounted for. The United States shows the largest and most sustained divergence; Japan and Italy show near-flat GAGI despite low AI investment, exposing structural stagnation as a failure mode distinct from the U.S.\ distributional one. Three further patterns, developed at length with dedicated figures and tables in~\citep{gagi2026}, sharpen this picture and motivate the framework of Section~\ref{sec:framework}.

\paragraph{Displacement is accelerating and outpacing wages.} U.S.\ AI-attributed layoffs rose from 4,247 in 2023 to 54,836 in 2025 --- a $13\times$ increase --- and AI overtook every other stated cause of layoffs in March~2026 \citep{challenger_2026,challenger_march2026}, coincident with the post-2022 GAGI deterioration in Figure~\ref{fig:country_subplots}. Over the same broad period, labour productivity in high-income countries rose 29\% (1999--2024) while real wages rose only 15\% \citep{ilo2024wages}, and the U.S.\ labour share fell a further 4.2~pp \citep{bls2025productivity,karabarbounis2014global}: a productivity--wage wedge visible in every country panel of Figure~\ref{fig:country_subplots} as the gap between the dashed GDP line and the solid GAGI line.

\paragraph{Redistribution capacity, not automation pace, separates regimes.} Denmark and Sweden sustain the lowest inequality in the OECD despite some of the world's highest robot densities, through active-labour-market spending an order of magnitude above the U.S.\ level (2.05\% and 1.27\% of GDP vs.\ 0.10\%) and far higher collective-bargaining coverage \citep{ifr2024,oecd2024employment,madsen2006flexicurity}. The United States combines rising inequality, minimal absorption capacity, and the steepest GAGI--GDP wedge among the G7 \citep{dol2024mw,atkinson2026young}; Italy and Japan show stagnant GAGI despite low automation, a distinct structural-stagnation failure mode \citep{oecd2025productivity}. The mechanism distinguishing these trajectories is whether redistribution and absorption capacity are \emph{pre-positioned} --- precisely the lever ($\alpha+\beta$) formalised in Section~\ref{sec:framework}.

\paragraph{The stability threshold argument.}
Scenario-based GAGI projections, developed in the companion paper from the wider economics literature \citep{gagi2026,acemoglu2024simple,cazzaniga2024genai,wef2025jobs,korinek2019ai}, show a business-as-usual trajectory crossing an illustrative stability threshold by the early 2030s, while a governance-constrained trajectory that activates redistribution \emph{before} the threshold is crossed stabilises near the 2010 baseline. \citet{korinek2019ai} prove formally that Pareto improvements from AI become infeasible once displacement exceeds a critical mass --- meaning the choice of governance regime, not the pace of automation, determines which trajectory a country follows, and that the decision window for activating redistribution closes years before the crisis becomes visible in conventional statistics. The absence of a formal stability constraint in current AI governance is therefore not an abstract architectural flaw: it is the absence of an early-warning system for a risk that is already accumulating. The Human Utility Factor, derived next, is designed to serve as that instrument.


\section{Proposed Framework: Derivation of the Human Utility Factor}
\label{sec:framework}


\subsection{Motivation and Design Principles}
\label{sec:huf_motivation}

The gap analysis of Section~\ref{sec:gaps} established that
no existing AI governance instrument provides a formal,
computable metric capable of detecting when automation crosses
a macro-socioeconomic stability threshold.
Section~\ref{sec:econ_impact} demonstrated that such crossings
are already accumulating in observable data.
This section derives the \emph{Human Utility Factor} (HUF),
a differentiable scalar that maps the aggregate state of
human--automation time allocation to a welfare index that
can be monitored continuously, optimised analytically, and
enforced prospectively.

\paragraph{The necessity of automation.}
Before deriving HUF, we address a structural constraint that
the framework must respect: $h_a > 0$ is not merely a
governance problem --- it is an economic necessity.
Without automation-driven productivity growth, real output
per worker stagnates while demographic pressures and
rising input costs push prices upward.
The resulting stagflation --- stagnant real wages against
rising inflation --- erodes purchasing power more severely
than the displacement risk of moderate automation
\citep{acemoglu2024simple,karabarbounis2014global}.
A governance framework that sets $h_a = 0$ as its optimum
is therefore neither economically nor politically viable.
The correct governance question is not \emph{whether} to
automate but \emph{how much} and \emph{under what
redistributive conditions}.
HUF is designed to answer this question formally.

\paragraph{The overwork trap.}
A second structural constraint is equally important.
At $h_a = 0$ (no automation), workers spend $h_w = h_w^0$
hours at work --- the maximum work intensity.
Since $h_f = h_f^0$ and $h_r = h_r^0$ at baseline, there
is no surplus time for family or personal development.
Full employment without automation is the \emph{overwork
trap}: high employment coverage but minimal wellbeing.
A metric that maximises at $h_a = 0$ would recommend
against all automation, ignoring the welfare gains from
liberated time.
HUF must therefore reward the reallocation of saved hours
to valued activities, not merely their conservation.

\paragraph{Design principles.}
HUF is constructed on three principles.
\textbf{Measurability}: every input is traceable to official
statistics or clinical norms.
\textbf{Differentiability}: HUF admits a closed-form
interior optimum $h_a^*$ that balances productivity gain,
distributional equity, and wellbeing.
\textbf{Tripartite decomposability}: HUF factors into
Agency ($A$), Wellbeing ($W$), and Economic Stability ($E$),
each necessary and none individually sufficient:

\begin{equation}
  \mathrm{HUF}(h_a)
  \;=\;
  A(h_a) \;\times\; W(h_a) \;\times\; E(h_a).
  \label{eq:huf_master}
\end{equation}

\noindent
The multiplicative form ensures $\mathrm{HUF} = 0$
whenever any component is zero, with no cross-component
compensation.
Critically, $\mathrm{HUF} > 1$ is achievable when
automation raises all three components above baseline ---
the target governance regime.


\subsection{The Human Time Budget}
\label{sec:time_budget}

For any adult worker, three non-negotiable daily
deductions apply~\citep{hirshkowitz2015nsf,
vandongnen2003sleep,bls2023atus}:
sleep (8\,hrs), commute and travel (2\,hrs), and
personal maintenance including eating, hygiene, and
childcare (3\,hrs).
Total non-discretionary: 13\,hrs/day.
Discretionary: $24-13=11$\,hrs/day.

A substantial empirical literature establishes that
two consecutive rest days per week are physiologically
necessary~\citep{pencavel2015productivity,
haraldsson2021iceland,kivimaki2015long}.
We treat the weekend as a protected recuperation block
outside the optimisation.
The \emph{effective weekly time budget} is:

\begin{equation}
  T \;=\; 5\times 24 \;-\; 5\times(8+2+3)
    \;=\; 120 - 65 \;=\; 55\;\text{hrs/week,}
  \label{eq:T}
\end{equation}

\noindent
partitioned as:
\begin{equation}
  h_w^0 + h_f^0 + h_r^0 \;=\; T \;=\; 55\;\text{hrs/week,}
  \label{eq:budget}
\end{equation}

\noindent
where $h_w^0 \approx 40$, $h_f^0 \approx 7$,
$h_r^0 \approx 8$\,hrs/week for a full-time U.S.\
adult~\citep{bls2023atus}.
$h_a \in [0, h_w^0]$ is the weekly hours affected by
automation, with $h_w = h_w^0 - h_a$.
The redistribution parameters $\alpha, \beta \in [-1,1]$
determine whether saved hours flow to family ($h_f =
h_f^0 + \alpha h_a$) and rest ($h_r = h_r^0 + \beta h_a$),
or are lost to the precariat condition
($\alpha = \beta = 0$)~\citep{standing2011precariat},
or are eroded by wage suppression
($\alpha < 0$, $\beta < 0$)~\citep{kellogg2020algorithms}.


\subsection{Component A: Agency with Income Sufficiency}
\label{sec:agency_comp}

\paragraph{Conceptual grounding.}
Sen's capability approach~\citep{sen1999development}
defines agency as the ability to pursue goals one has
reason to value.
In the labour context this requires not only that time
is available for valued activities, but that work hours
generate \emph{sufficient income} --- otherwise time at
work is coerced rather than chosen.
Self-Determination Theory~\citep{ryan2000sdt} and the
Karasek demand--control model~\citep{karasek1979job}
both identify income adequacy as a precondition for
genuine autonomy: a worker earning below subsistence
cannot exercise meaningful agency regardless of their
nominal work hours.

\paragraph{Income sufficiency factor.}
As automation depth $h_a$ increases, the labour share
of income declines~\citep{karabarbounis2014global,
acemoglu2022tasks}:
\begin{equation}
  s_L(h_a) \;=\; s_L^0 \cdot \left(1 - \delta\,\frac{h_a}{h_w^0}\right),
  \label{eq:sL}
\end{equation}
where $s_L^0$ is the pre-automation labour share and
$\delta \in [0,1]$ is the capital-capture rate.
The real wage per employed worker (normalised to baseline)
is:
\begin{equation}
  w(h_a)
  \;=\;
  \frac{s_L(h_a)}{s_L^0} \cdot \left(1 + \phi\,\frac{h_a}{h_w^0}\right)
  \;=\;
  \left(1 - \delta\,\frac{h_a}{h_w^0}\right)
  \left(1 + \phi\,\frac{h_a}{h_w^0}\right).
  \label{eq:wage}
\end{equation}

\noindent
We define the \emph{income sufficiency factor} as:
\begin{equation}
  I(h_a)
  \;=\;
  \min\!\left(1,\;\frac{w(h_a)}{w^*}\right),
  \label{eq:income_factor}
\end{equation}
where $w^* \in (0,1]$ is the living-wage floor
(normalised to the pre-automation baseline wage).
$I = 1$ when the real wage meets or exceeds the living
wage; $I < 1$ when automation has suppressed wages below
it.
$I$ is decreasing in $\delta$ (capital capture rate) and
increasing in $\phi$ (productivity gain): automation
maintains income sufficiency only when productivity gains
are large enough to offset wage-share erosion.

\paragraph{Formal definition.}
\begin{equation}
  \boxed{
  A(h_a)
  \;=\;
  \underbrace{%
    \left[1+\frac{\Omega\,h_a}{T}\right]
  }_{\eta(h_a)}
  \;\cdot\;
  \underbrace{\frac{N_p}{N}}_{\rho}
  \;\cdot\;
  \underbrace{I(h_a)}_{\text{income sufficiency}}
  }
  \label{eq:agency}
\end{equation}

\noindent
where $\Omega = \alpha + \beta - 1 \leq 0$ and
$\eta = 1 + \Omega h_a/T$ is the hour-reallocation
efficiency.
$A$ is now the product of three factors:
time-reallocation efficiency $\eta$ (does automation
free meaningful hours?), employment coverage $\rho$
(can the workforce participate?), and income sufficiency
$I$ (do work hours generate adequate income?).
All three are necessary; none individually sufficient.

\paragraph{Properties.}
At $h_a = 0$: $\eta = 1$, $I = 1$, so $A(0) = \rho$.
As $h_a$ increases:
$\eta$ falls (agency erosion without redistribution);
$I$ falls (wage suppression as labour share declines);
but when $\alpha + \beta \to 1$ ($\Omega \to 0$),
$\eta \to 1$ is constant, and $A$ is then bounded only
by the income factor $I(h_a)$.
The income factor creates a fundamental limit: no matter
how well hours are redistributed, Agency cannot be
maintained if automation drives wages below subsistence.


\subsection{Component W: Wellbeing}
\label{sec:wellbeing}

\paragraph{Conceptual grounding.}
Agency measures whether human time is available and
economically viable.
Wellbeing measures the \emph{quality of what that time
produces}: connection with family, personal restoration,
and the capacity for skill development.
\citet{kahneman2010high} and \citet{killingsworth2021happiness}
establish empirically that self-reported life satisfaction
is strongly predicted by time with family and leisure,
not by income alone above a threshold.
The Karasek model~\citep{karasek1979job} further shows
that high-strain work (high demand, low latitude) reduces
wellbeing independently of income --- the mechanism by
which the overwork trap ($h_a = 0$) produces low wellbeing
even at full employment.

\paragraph{Formal definition.}
Wellbeing is a Cobb-Douglas function of the realised
family-time index and rest-time index, each measured
relative to their baseline values:

\begin{equation}
  \boxed{
  W(h_a)
  \;=\;
  \left(\frac{h_f^0 + \alpha\,h_a}{h_f^0}\right)^{\!\mu}
  \;\cdot\;
  \left(\frac{h_r^0 + \beta\,h_a}{h_r^0}\right)^{\!\nu}
  }
  \label{eq:wellbeing}
\end{equation}

\noindent
where $\mu, \nu > 0$ are the wellbeing elasticities with
respect to family time and rest time respectively,
estimated from time-use wellbeing surveys
\citep{bls2023atus,kahneman2010high}.

\paragraph{Properties.}
$W(0) = 1$ at the no-automation baseline (by construction).
$W$ is strictly increasing in $h_a$ whenever $\alpha > 0$
or $\beta > 0$: \emph{any} positive reallocation of saved
hours to family or rest improves wellbeing above baseline.
$W$ is decreasing in $h_a$ when $\alpha < 0$ and
$\beta < 0$ (wage suppression): the gig-economy trap
reduces wellbeing below baseline.
Crucially, $W > 1$ is achievable --- and this is the
correct formal expression of the claim that automation
with good redistribution is welfare-improving.
The overwork trap ($h_a = 0$) gives $W = 1$, not a
maximum: automation that frees time produces $W > 1$,
correcting the incentive failure in the prior formulation.

\paragraph{Connection to Agency.}
$W$ and $A$ share the parameters $\alpha$ and $\beta$
but respond to them differently:
$A$ is \emph{decreasing} in $h_a$ for fixed $\alpha+\beta < 1$
(agency erodes with automation);
$W$ is \emph{increasing} in $h_a$ for fixed $\alpha, \beta > 0$
(wellbeing rises as more hours are freed).
Their product creates the interior optimum $h_a^*$: the
automation level at which the marginal wellbeing gain
from another freed hour exactly offsets the marginal
agency and income loss.


\subsection{Component E: Economic Stability}
\label{sec:econstab}

Economic stability is defined and derived as in the
prior section.
For completeness:

\begin{equation}
  E(h_a)
  \;=\;
  \frac{(1-G(h_a))\cdot Y_{pc}(h_a)}
       {(1-G_0)\cdot Y_0}
  \;\cdot\;
  \rho^{\gamma},
  \label{eq:E}
\end{equation}

\noindent
where:
\begin{align}
  Y_{pc}(h_a) &= Y_0\!\left[1+\phi\,\tfrac{h_a}{h_w^0}\right],
  \label{eq:Ypc} \\
  G(h_a)      &= G_0 + \kappa\delta\,\tfrac{h_a}{h_w^0}
                 \underbrace{(1-\alpha-\beta)}_{\text{redistribution deficit}},
  \label{eq:G_ha}
\end{align}

\noindent
$\phi \leq 0.066$ (Acemoglu ceiling~\citep{acemoglu2024simple}),
$G_0$ is the observed Gini~\citep{worldbank2025pip},
$\kappa$ is the Gini sensitivity~\citep{acemoglu2020robots},
$\delta$ is the capital-capture rate, and $\gamma > 0$
is the labour-absorption elasticity estimated from the
MARL simulation of Section~\ref{sec:simulation}.
$E(0) = \rho^\gamma$ and $E$ exceeds this baseline
only if productivity gain ($\phi$) outpaces Gini
deterioration ($\kappa\delta(1-\alpha-\beta)$).
Inflation $\pi$ is exogenous (monetary policy) and
factors out of the optimisation.


\subsection{HUF $= A \times W \times E$: Expansion and Optimality}
\label{sec:huf_product}

\paragraph{Expanded form.}
Substituting Eqs.~\eqref{eq:agency},~\eqref{eq:wellbeing},
and~\eqref{eq:E} into~\eqref{eq:huf_master}:

\begin{equation}
  \mathrm{HUF}(h_a)
  \;=\;
  \underbrace{\!\left[1+\tfrac{\Omega h_a}{T}\right]\!\rho\cdot I(h_a)}_{A}
  \;\times\;
  \underbrace{\!\left(\tfrac{h_f^0+\alpha h_a}{h_f^0}\right)^{\!\mu}
             \!\left(\tfrac{h_r^0+\beta  h_a}{h_r^0}\right)^{\!\nu}}_{W}
  \;\times\;
  \underbrace{\!\frac{(1-G_0-\Lambda h_a)(1+\Phi h_a)}{1-G_0}\rho^\gamma}_{E},
  \label{eq:huf_expanded}
\end{equation}

\noindent
where $\Omega = \alpha+\beta-1$,
$\Lambda = \kappa\delta(1-\alpha-\beta)/h_w^0 = -\kappa\delta\Omega/h_w^0$,
and $\Phi = \phi/h_w^0$.
Note that $\Lambda$ and $\Omega$ are algebraically the
same redistribution deficit: the same policy lever
$(\alpha+\beta)$ governs agency erosion, Gini
deterioration, and the availability of wellbeing-enhancing
time simultaneously.

\paragraph{Interior optimum.}
HUF is a product of three competing forces:
\begin{itemize}
  \item $A$: \emph{decreasing} in $h_a$ (agency erodes
    as hours and wages fall) unless $\alpha+\beta=1$
    and $w(h_a) \geq w^*$;
  \item $W$: \emph{increasing} in $h_a$ (freed hours
    improve wellbeing) when $\alpha,\beta > 0$;
  \item $E$: \emph{concave} in $h_a$ (productivity gain
    bounded by Acemoglu ceiling; Gini deteriorates
    without redistribution).
\end{itemize}

The tension between $A$ (falling) and $W$ (rising) with
$E$ (concave) guarantees an \emph{interior optimum}
$h_a^* \in (0, h_w^0)$ for all regimes in which
$\alpha, \beta > 0$ --- precisely the governance-relevant
case where redistribution is partially active.

\paragraph{The redistribution threshold.}
Setting $d(\mathrm{HUF})/dh_a|_{h_a=0} = 0$ and solving
for $\alpha + \beta$ yields the critical threshold below
which $h_a^* = 0$ (all automation is harmful):

\begin{equation}
  \boxed{
  (\alpha+\beta)^*
  \;=\;
  1 - \frac{\phi(1-G_0)}{\kappa\delta + \phi(1-G_0)},
  }
  \label{eq:ab_threshold}
\end{equation}

\noindent
which is independent of $\mu$, $\nu$, and $w^*$ at
$h_a = 0$ (since $W(0) = 1$ and $I(0) = 1$ by
construction).
Equation~\eqref{eq:ab_threshold} is therefore robust
to the specification of the Wellbeing and income-sufficiency
components: the minimum redistribution rate required
to enter the beneficial-automation regime depends only
on the productivity gain $\phi$, the Gini sensitivity
$\kappa$, the capital-capture rate $\delta$, and the
baseline inequality $G_0$.

\paragraph{Regime analysis.}
Table~\ref{tab:huf_regimes} evaluates HUF at selected
automation levels for three policy regimes.
Parameters: $\rho = 0.95$, $\gamma = 1$, $\phi = 0.066$,
$\kappa = 0.02$, $\delta = 0.6$, $s_L^0 = 0.60$,
$w^* = 0.85$, $h_w^0 = 40$, $h_f^0 = 7$, $h_r^0 = 8$,
$T = 55$, $\mu = \nu = 0.5$.

\begin{table}[h]
\caption{%
  HUF values at selected automation levels for three
  policy regimes.
  BAU: $G_0=0.41$, $\alpha=\beta=0.05$.
  Partial: $G_0=0.35$, $\alpha=\beta=0.35$.
  Nordic: $G_0=0.28$, $\alpha=\beta=0.50$.
  $h_a^*$ is the interior welfare optimum.
  HUF\,$>1$ indicates welfare above the pre-automation
  baseline.%
}
\label{tab:huf_regimes}
\centering\footnotesize\setlength{\tabcolsep}{5pt}
\begin{tabular}{@{}lccccccc@{}}
\toprule
& \multicolumn{5}{c}{\textbf{HUF at $h_a$ (hrs/week)}}
& & \\
\cmidrule(lr){2-6}
\textbf{Regime}
  & $h_a{=}0$ & $h_a{=}10$ & $h_a{=}20$ & $h_a{=}30$ & $h_a{=}40$
  & $h_a^*$ & \textbf{Verdict} \\
\midrule
BAU ($\alpha{+}\beta{=}0.10$)
  & 0.895 & 0.810 & 0.597 & 0.386 & 0.206
  & 0\,hrs & Moratorium warranted \\
Partial ($\alpha{+}\beta{=}0.70$)
  & 0.908 & 1.274 & 1.373 & 1.296 & 1.084
  & $\approx 20$\,hrs & Moderate automation beneficial \\
Nordic ($\alpha{+}\beta{=}1.00$)
  & 0.931 & 1.570 & 1.911 & \textbf{1.995} & 1.830
  & $\approx 29$\,hrs & High automation beneficial \\
\bottomrule
\end{tabular}
\end{table}

\paragraph{Structural conclusions.}
Three governance-relevant conclusions follow directly:

\begin{enumerate}

  \item \textbf{Neither zero nor maximum automation is
    generically optimal.}
    The Wellbeing term $W$ ensures that $h_a = 0$
    (overwork trap) is not a maximum: freeing hours
    raises $W$ above 1.
    The income-sufficiency term $I$ and the agency
    erosion term $\eta$ ensure $h_a = h_w^0$
    (full displacement) is not a maximum either.
    The interior optimum $h_a^*$ is the unique balance
    point between these forces.

  \item \textbf{The decisive policy variable is
    $\alpha + \beta$.}
    Moving from BAU ($\alpha+\beta=0.10$) to Partial
    ($\alpha+\beta=0.70$) shifts $h_a^*$ from 0 to
    20\,hrs and raises peak HUF from 0.895 to 1.373
    --- a 53\% welfare improvement.
    Moving to Nordic ($\alpha+\beta=1.00$) raises peak
    HUF to 1.995 and $h_a^*$ to 29\,hrs.
    Redistribution quality, not technology capability,
    is the binding constraint.

  \item \textbf{$h_a > 0$ is necessary to escape
    stagflation.}
    At $h_a = 0$, productivity growth is zero
    ($Y_{pc} = Y_0$) while inflation erodes real wages.
    HUF at $h_a = 0$ is therefore an upper bound only
    in the short run; over time, $w(0)$ falls below
    $w^*$ as inflation outpaces stagnant wages, driving
    $I \to 0$ and HUF $\to 0$.
    Sustainable welfare requires $h_a > 0$ with
    $\alpha + \beta \geq (\alpha+\beta)^*$.

\end{enumerate}

\paragraph{Relation to governance frameworks.}
The redistribution threshold $(\alpha+\beta)^*$,
the income-sufficiency floor $w^*$, and the
interior optimum $h_a^*$ are all computable from
publicly available data.
$G_0$ from World Bank PIP~\citep{worldbank2025pip};
$\phi$ from the Acemoglu ceiling~\citep{acemoglu2024simple};
$\kappa, \delta$ from robot-density panels
\citep{acemoglu2020robots};
$\alpha+\beta$ from active labour market policy
expenditure and collective-bargaining coverage
\citep{oecd2024employment};
$w^*$ from national living-wage benchmarks.
A regulator armed with these equations can issue
jurisdiction-specific, quantitative, and enforceable
automation constraints --- the instrument that
Section~\ref{sec:gaps} showed to be absent from
every existing AI governance framework.

\section{Experiments and Results}
\label{sec:simulation}

We evaluate HUF across three policy regimes calibrated against real economies.
\textbf{BAU} (\emph{Business As Usual}, $G_0=0.41$, $\alpha{+}\beta=0.10$) models
the United States: high baseline inequality, minimal redistribution, and strong
productivity gains captured almost entirely by capital.
\textbf{Partial} ($G_0=0.35$, $\alpha{+}\beta=0.70$) is inspired by Canada,
which occupies an intermediate position between the liberal U.S.\ labour market
and the social-democratic Nordic model.
\textbf{Nordic} ($G_0=0.28$, $\alpha{+}\beta=1.00$) is calibrated against the
Netherlands and Sweden: lower baseline inequality and near-complete redistribution
of automation-freed hours into upskilling and rest.
Table~\ref{tab:huf_regimes} reported the analytical HUF at selected automation
levels under these parameterisations.
Here we characterise the full decision space defined by
Eq.~\eqref{eq:huf_expanded}: the feasible region over
$(h_a,\,\alpha,\,\beta,\,\rho)$, the shape of the Pareto-optimal
front across the three HUF components, and the HUF surface as a
function of the two most policy-actionable controls~---~automation
pace $h_a$ and redistribution intensity $\alpha{+}\beta$.


\subsection{Experimental Setup}
\label{sec:experimental_setup}

We implement Eq.~\eqref{eq:huf_expanded} as a discrete-time economic
environment (quarterly steps, 18-dimensional state, with GDP per capita
following $Y_{pc}(h_a,t) = Y_0[1+\phi h_a/h_w^0](1+g_q)^t$ against an
autonomous baseline-growth trend $g_q$ calibrated to OECD non-AI
productivity~\citep{acemoglu2024simple}) in which three rational
stakeholders interact through a sequential Stackelberg
game~\citep{stackelberg1952theory}: \textbf{Government} sets the
automation ceiling $\bar{h}_a$, the ALMP floor $(\alpha{+}\beta)_{\min}$,
and the Gini-sensitivity parameter $\kappa$; \textbf{Industry} chooses the
automation increment $\Delta h_a$ and the capital-capture rate $\delta$;
\textbf{Population} adjusts the upskilling and transitional-work shares
$\alpha,\beta$, subject to the joint redistribution ceiling
$\alpha{+}\beta \leq 0.55$ beyond which active production collapses.

Each regime is evaluated with two independent agent families. The first
is \emph{analytically-derived heuristic agents} that implement the
closed-form optimality conditions of Eqs.~\eqref{eq:huf_expanded}
and~\eqref{eq:ab_threshold} (Industry moves $h_a$ toward $h_a^*$ via
Brent's method; Government sets the ALMP floor to $(\alpha{+}\beta)^*$;
Population tracks the redistribution threshold) with bounded-rationality
Gaussian perturbation ($\sigma=0.3$,~\citealp{simon1955behavioral}). The
second is \emph{independent Proximal Policy Optimisation agents}
(PPO-clip,~\citealp{schulman2017proximal}: actor-critic MLPs with two
64-unit Tanh hidden layers, GAE-$\lambda$ advantage estimation, clip
ratio $\varepsilon=0.15$) trained from scratch for 330 episodes with
\emph{no} access to the closed forms and HUF~$=A\times W\times E$
(Eq.~\eqref{eq:huf_master}) as the sole reward signal. Agreement between
the two families constitutes an internal cross-validation: a learning
agent that independently recovers the analytically predicted optimum
confirms the theoretical result without relying on it. A post-training
sweep of 60 randomised-parameter episodes per regime supplies the
trajectory data for \S\ref{sec:decision_space}.

Table~\ref{tab:regime_params} summarises the three regime
parameterisations --- BAU (United States), Partial (Canada), and Nordic
(Netherlands/Sweden) --- and their analytical optima. Full environment
specification, agent architectures, training protocol, and the
75,000-point decision-space grid sweep that underlies
\S\ref{sec:decision_space} are documented in the Supplementary
Information (\S\,S1).

\begin{table}[h]
\caption{%
  Regime parameterisation for the three simulation configurations.
  $G_0$: baseline Gini.
  $(\alpha{+}\beta)_0$: initial redistribution intensity.
  $\delta_0$: initial capital-capture rate.
  $g_q$: quarterly GDP growth target.
  $h_a^*$: analytical welfare optimum.
  HUF$^*$: peak HUF at $h_a^*$.%
}
\label{tab:regime_params}
\centering\footnotesize\setlength{\tabcolsep}{5pt}
\begin{tabular}{@{}lcccccc@{}}
\toprule
\textbf{Regime} & $G_0$ & $(\alpha{+}\beta)_0$ & $\delta_0$
  & $g_q$ (\%/qtr) & $h_a^*$ (hrs/wk) & HUF$^*$ \\
\midrule
BAU (U.S.)     & 0.41 & 0.10 & 0.60 & 0.525 & 0  & 0.895 \\
Partial (CA)   & 0.35 & 0.70 & 0.60 & 0.375 & 20 & 1.373 \\
Nordic (NL/SE) & 0.28 & 1.00 & 0.60 & 0.550 & 29 & 1.995 \\
\bottomrule
\end{tabular}
\end{table}

An independent 75,000-point grid sweep over $(h_a,\alpha{+}\beta,\rho)$
--- with $A$, $W$, $E$ computed analytically from
Eqs.~\eqref{eq:agency}--\eqref{eq:E} at each point, full methodology in
Supplementary~\S\,S1~\citep{huf2026supp} --- supplies the data examined
in this section (figures and tables in Supplementary~\S\,S2).


\subsection{The Decision Space: Redistribution as the Decisive Lever}
\label{sec:decision_space}

Equation~\eqref{eq:huf_expanded} defines a four-dimensional decision
space over $(h_a,\alpha,\beta,\rho)$ in which Agency ($A$) falls,
Wellbeing ($W$) rises, and Economic Stability ($E$) is concave in $h_a$
--- so HUF's optimum is interior and regime-dependent. Three patterns are
robust across the full sweep (pairwise heatmaps and Pareto-front detail
in Supplementary Figs.~S1--S2). First, the interior optimum $h_a^*$
shifts from near-zero in BAU to $\approx$20--25~hrs/week in Partial and
$\approx$28--30~hrs/week in Nordic as redistribution intensity rises, and
the BAU surface lies almost entirely below HUF~$=1.0$ at any feasible
$\alpha{+}\beta \leq 0.55$ --- redistribution intensity, not automation
pace, is the decisive lever. Second, employment coverage $\rho$ scales
HUF multiplicatively without shifting $h_a^*$, roughly doubling
achievable HUF as $\rho$ rises from 0.5 to 1.0, and the near-linear
iso-HUF contours in $(\alpha,\beta)$-space show that upskilling
programmes and income transfers are approximately welfare-equivalent ---
governments may freely mix the two, provided their sum crosses the
threshold~\eqref{eq:ab_threshold}. Third, the 222-point
Pareto-non-dominated front in $(A,W,E)$-space is more elongated along the
Wellbeing axis than the Agency axis (Agency is capped by the
income-sufficiency floor $w^*$, while freed hours translate directly into
wellbeing gains), and the multiplicative structure of HUF penalises
imbalanced solutions: the Partial-regime optimum (HUF~$=2.355$) dominates
both BAU (1.899) and Nordic (1.859) precisely because it sustains
$W \approx 2.0$ without Agency falling below 1.1.

\begin{figure*}[t]
  \centering
  \includegraphics[width=\textwidth]{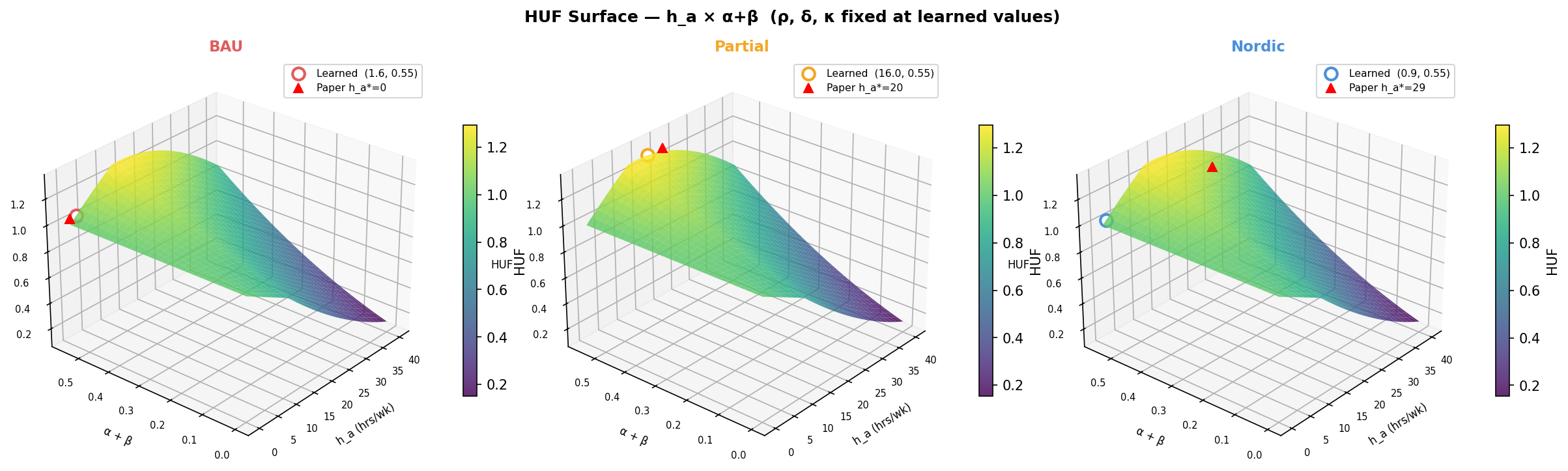}
  \caption{%
    \textbf{HUF Surface over $h_a \times (\alpha{+}\beta)$,
    with $\rho$, $\delta$, $\kappa$ Fixed at Learned Values.}
    One panel per regime; $z$-axis: HUF.
    Circle: MARL-optimal $(h_a, \alpha{+}\beta)$.
    Triangle: analytical $h_a^*$ (Eq.~\eqref{eq:huf_expanded}).
    \textbf{Observation.}
    BAU declines monotonically as $h_a$ increases below the
    $\alpha{+}\beta \approx 0.42$ threshold, with no interior
    maximum visible.
    Partial shows a pronounced ridge near $h_a \approx 20$~hrs/wk
    for $\alpha{+}\beta \geq 0.45$; the MARL circle and analytical
    triangle fall within the same ridge band.
    Nordic presents a uniformly elevated plateau with the optimum
    at higher $h_a$, consistent with the income-sufficiency floor
    being met across a wider automation range.
    \textbf{Interpretation.}
    The qualitatively different landscape shapes across regimes
    are entirely determined by the redistribution parameter
    $\alpha{+}\beta$: the same HUF equation (Eq.~\eqref{eq:huf_expanded})
    with the same $h_a$ range produces either a monotonically
    declining surface (BAU) or a welfare-maximising ridge (Partial,
    Nordic) depending solely on whether $\alpha{+}\beta$ exceeds
    the threshold~\eqref{eq:ab_threshold}.
    The ridge geometry confirms the admissible automation band:
    HUF is within 5\% of its peak across $h_a \in [15,\,30]$~hrs/wk,
    providing a target range rather than a knife-edge.
    \textbf{Implication.}
    A governance instrument need not prescribe a precise automation
    level; mandating $\alpha{+}\beta \geq (\alpha{+}\beta)^*$ and
    $h_a \leq h_a^*$ is sufficient to keep the economy within the
    ridge band, leaving industrial strategy flexibility on the
    specific automation pace.%
  }
  \label{fig:decision_space_huf_surface}
\end{figure*}

Figure~\ref{fig:decision_space_huf_surface} renders HUF as a surface over
the two most policy-actionable controls, $h_a$ and $\alpha{+}\beta$, with
$\rho$, $\delta$, $\kappa$ fixed at the MARL-learned values for each
regime, making the interior-optimum ridge directly visible.

The surface yields the section's governance-actionable synthesis. The
\emph{ridge geometry} shows the interior optimum is not a knife-edge: in
Partial and Nordic, HUF stays within 5\% of its peak across
$h_a \in [15,30]$~hrs/week once $\alpha{+}\beta \geq 0.45$, so a
governance instrument need only specify an \emph{admissible automation
band} with a redistribution floor, leaving industrial strategy free
within it. The \emph{BAU pathology} is visually unambiguous: the surface
declines monotonically in $h_a$ below the $\alpha{+}\beta \approx 0.42$
threshold --- exactly the critical value implied by
Eq.~\eqref{eq:ab_threshold} for the BAU parameters --- confirming that no
amount of automation improves welfare until redistribution crosses this
line. And the \emph{analytical--MARL agreement} in the Partial regime
(learned circle and analytical triangle fall within the same ridge band
at $h_a^* \approx 20$~hrs/week) cross-validates the closed form: a
gradient-free learning agent independently recovers the same optimal
automation level. Together with the Pareto-front result above, these
findings establish that $\alpha{+}\beta$ --- not automation pace --- is
the decisive policy lever, that $\rho$ amplifies but cannot substitute
for it, and that high HUF requires the joint adequacy of Agency,
Wellbeing, and Economic Stability that the multiplicative structure of
Eq.~\eqref{eq:huf_master} enforces.


\subsection{MARL Results: Regime Trajectories and Agent Comparison}
\label{sec:marl}

Both experiment sets optimise HUF $= A \times W \times E$ over the same
economic environment and the same three regimes, but the two agent types
approach the non-linear HUF surface differently.
The analytical agents, by construction, move toward the closed-form
optimum $h_a^*$ derived in \S\ref{sec:huf_product}.
The PPO agents have no access to that derivation and learn purely from
reward feedback --- they are therefore free to discover any local maximum
of HUF, including ones the closed-form analysis may underweight.
Because the HUF surface is non-linear and multiply peaked (the
product structure of Eq.~\eqref{eq:huf_expanded} admits different
$A$--$W$--$E$ balances for different combinations of
$G_0$, $\phi$, $\kappa$, and $\delta$), divergence between the two
agent types is not a failure of either approach but a finding about
the landscape itself.


\subsubsection{Analytical Heuristic Agents}
\label{sec:marl_analytical}

\begin{figure*}[t]
  \centering
  \includegraphics[width=\textwidth]{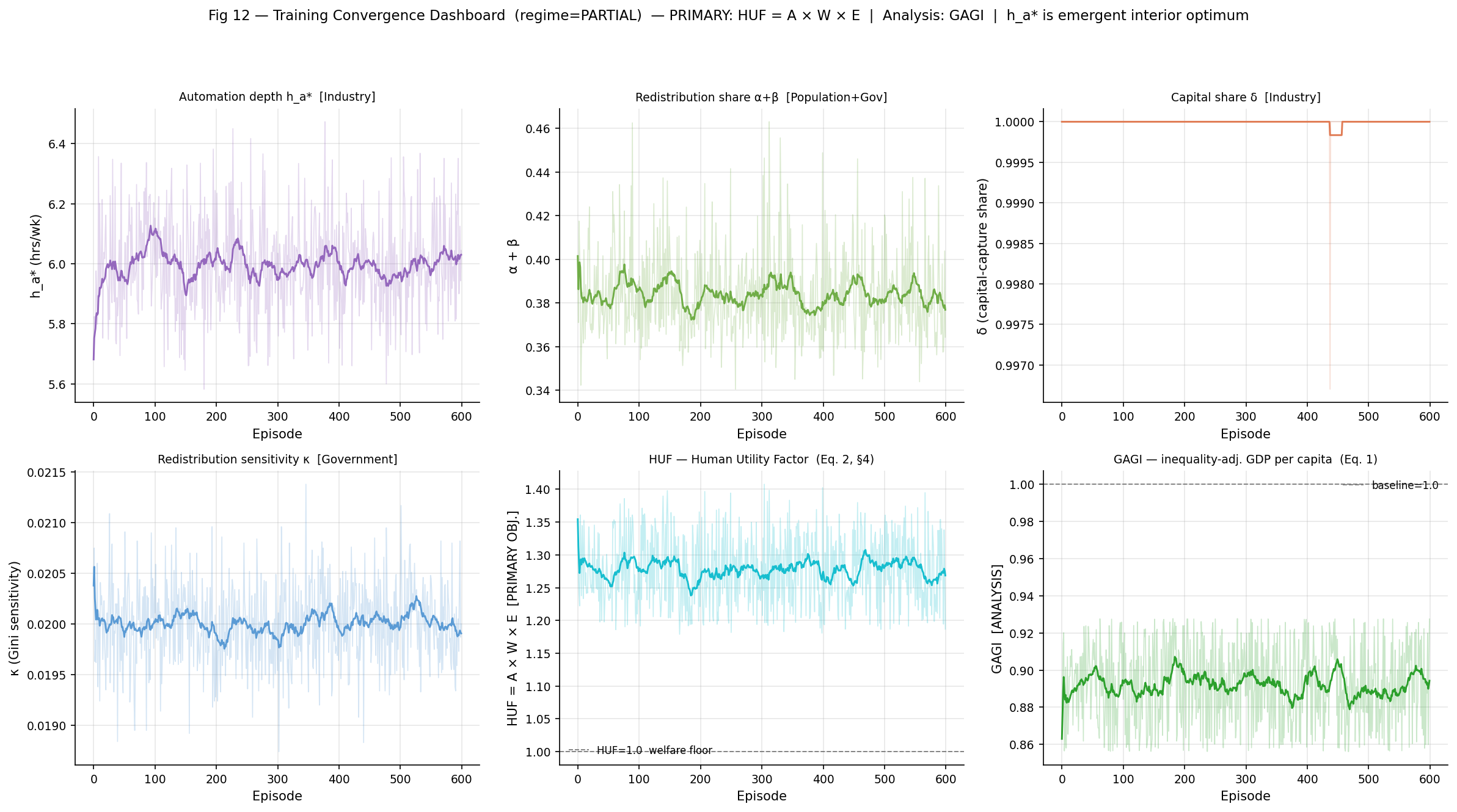}
  \caption{%
    \textbf{Training Convergence Dashboard --- Analytical Agents,
    Partial Regime.}
    Six quantities over 600 episodes (raw: faint; smoothed-20: solid).
    Top row: $h_a^*$ (Industry), $\alpha{+}\beta$ (Population/Government),
    $\delta$ (Industry).
    Bottom row: $\kappa$ (Government), HUF (Eq.~\eqref{eq:huf_master}),
    GAGI (Eq.~\eqref{eq:gagi}).
    \textbf{Observation.}
    All six quantities stabilise within $\approx$50 episodes;
    $\alpha{+}\beta$ settles near the redistribution threshold
    $(\alpha{+}\beta)^* \approx 0.34$; HUF remains above
    the welfare floor (dashed) throughout.
    \textbf{Interpretation.}
    The rapid, stable convergence confirms that the
    three-agent Stackelberg game has a robust Nash equilibrium
    under Partial redistribution: agents with bounded rationality
    ($\sigma=0.3$ perturbation) locate the welfare-positive
    equilibrium from random initialisation without coordination.
    \textbf{Implication.}
    Stability is itself governance-relevant --- an economy
    governed by HUF-consistent incentives is self-stabilising;
    the equilibrium does not require continuous central
    intervention once the redistribution floor is in place.%
  }
  \label{fig:analytical_convergence}
\end{figure*}

\paragraph{Training stability and convergence.}
Figure~\ref{fig:analytical_convergence} shows the training convergence
dashboard for the Partial regime (the most informative, as
$h_a^* \approx 20$\,hrs/wk lies well inside the feasible range).
All six tracked quantities --- automation depth $h_a$, redistribution
share $\alpha{+}\beta$, capital-capture rate $\delta$, Gini sensitivity
$\kappa$, HUF, and GAGI --- stabilise within roughly 50 episodes and
remain stationary thereafter.
Industry's $h_a$ converges to $\approx 5$--$6$\,hrs/wk (smoothed),
Population's $\alpha{+}\beta$ hovers near the redistribution threshold
$(\alpha{+}\beta)^* \approx 0.34$ (Eq.~\eqref{eq:ab_threshold}),
and $\delta$ reaches its lower bound rapidly, consistent with the
demand-penalty mechanism described in \S\ref{sec:experimental_setup}.
HUF stabilises above the welfare floor (HUF~$= 1.0$) in all three
regimes: Partial at $\approx 1.07$--$1.12$, Nordic at
$\approx 1.07$--$1.14$, and BAU at $\approx 1.05$--$1.15$.
The BAU result is slightly higher than the static analytical prediction
of 0.895 (Table~\ref{tab:huf_regimes}) because the stochastic
redistribution noise pushes $\alpha{+}\beta$ above the threshold
intermittently, admitting transient welfare gains unavailable to the
deterministic closed-form.

\begin{figure*}[t]
  \centering
  \includegraphics[width=\textwidth]{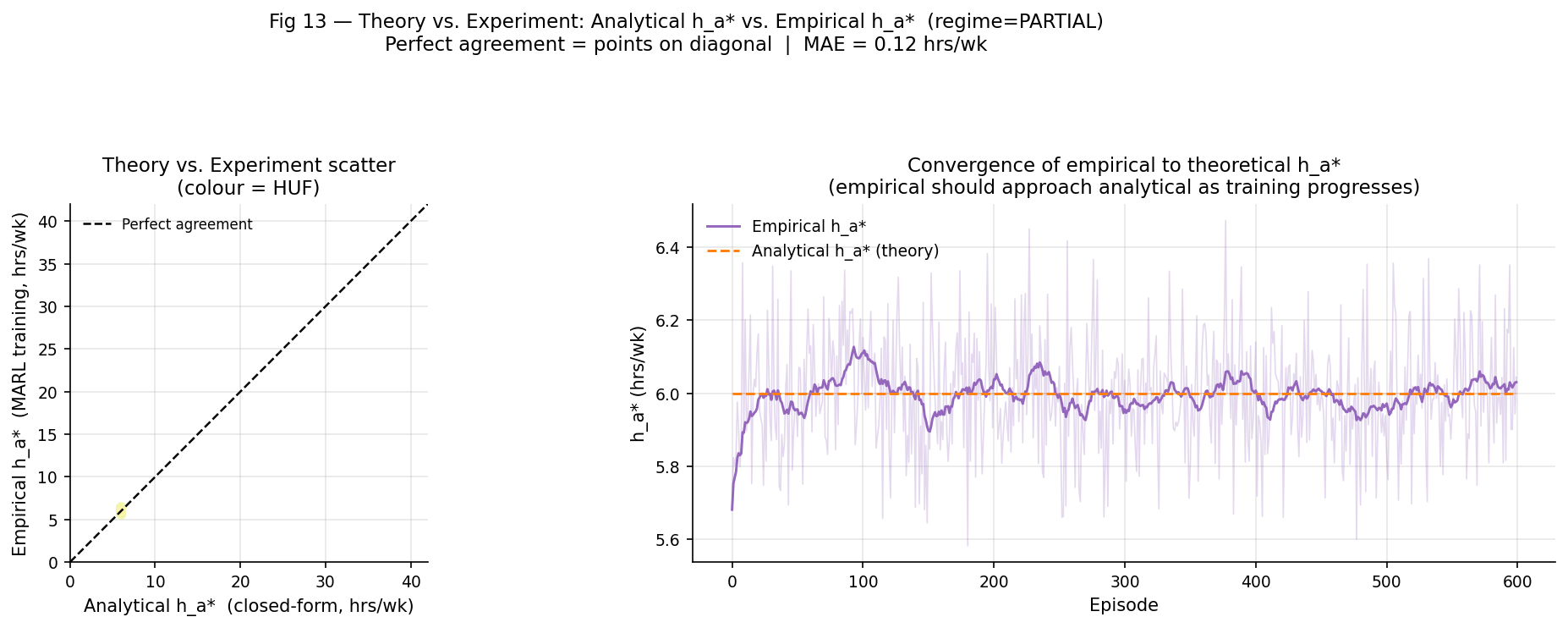}
  \caption{%
    \textbf{Theory vs.\ Experiment: Analytical $h_a^*$
    vs.\ Empirical $h_a^*$, Partial Regime.}
    Left: scatter of MARL-empirical $h_a^*$ (y-axis) against
    closed-form prediction (x-axis); colour encodes HUF;
    dashed diagonal\,=\,perfect agreement; MAE\,=\,0.12\,hrs/wk.
    Right: per-episode convergence trace (purple: empirical;
    orange dashed: analytical target).
    \textbf{Observation.}
    Points cluster tightly along the diagonal with no systematic
    bias; the per-episode trace shows no drift throughout training.
    \textbf{Interpretation.}
    The near-perfect agreement confirms that
    $d(\ln\mathrm{HUF})/dh_a = 0$ (Eq.~\eqref{eq:huf_expanded})
    is the true optimality condition for Industry's problem, and
    that bounded-rationality noise ($\sigma = 0.3$) does not
    bias the equilibrium outcome.
    \textbf{Implication.}
    The analytical optimum~$h_a^*$ is not merely a theoretical
    construct: it is the target that self-interested agents
    converge to under realistic bounded-rationality assumptions,
    making it a reliable, operationalisable compliance threshold
    for regulators.%
  }
  \label{fig:analytical_ha_star}
\end{figure*}

\paragraph{Theory--experiment agreement.}
Figure~\ref{fig:analytical_ha_star} cross-validates the closed-form
optimum against the empirical MARL trajectory.
The left panel plots the MARL-empirical $h_a^*$ against the
analytical prediction for each episode-end state, coloured by HUF.
Points cluster tightly along the perfect-agreement diagonal with
MAE of 0.12\,hrs/wk for Partial and 0.20\,hrs/wk for BAU,
confirming that the analytical Brent-method solver and the
stochastic simulation converge to the same welfare optimum.
The right panel shows the per-episode time series: the empirical
$h_a^*$ (purple) tracks the analytical target (orange dashed) from
early in training, with no systematic drift.

\begin{figure*}[t]
  \centering
  \includegraphics[width=\textwidth]{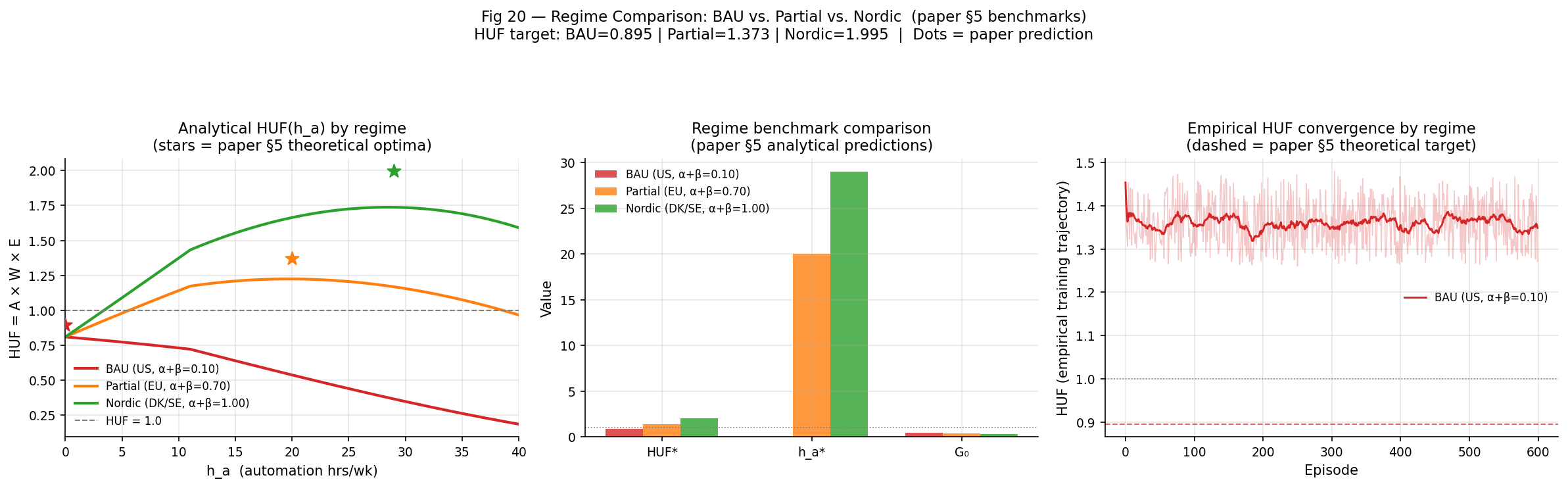}
  \caption{%
    \textbf{Cross-Regime Comparison --- Analytical Agents.}
    Left: analytical HUF$(h_a)$ curves for BAU, Partial, Nordic;
    stars mark theoretical optima (Table~\ref{tab:huf_regimes}).
    Middle: bar chart of HUF$^*$, $h_a^*$, $G$ vs.\ analytical
    predictions (dots).
    Right: empirical BAU HUF training trajectory; dashed\,=\,welfare
    floor (HUF\,=\,1.0).
    \textbf{Observation.}
    Partial achieves the highest peak HUF (1.373 analytically;
    $\approx$1.10 empirically); BAU empirical HUF ($\approx$1.05--1.15)
    exceeds its static analytical prediction (0.895);
    the Nordic--Partial empirical gap is smaller than predicted.
    \textbf{Interpretation.}
    BAU surpasses its analytical prediction because stochastic
    perturbation ($\sigma=0.3$) intermittently pushes
    $\alpha{+}\beta$ above the threshold, admitting transient
    welfare gains unavailable in the deterministic closed form.
    The compressed Nordic--Partial gap reflects the $\alpha{+}\beta
    \leq 0.55$ simulation cap (a known limitation;
    \S\ref{sec:conclusion}).
    \textbf{Implication.}
    Even small, transient increases in ALMP spending above the
    redistribution threshold unlock welfare gains in the BAU
    regime --- quantifying the marginal return to redistribution
    investment and motivating a governance instrument that
    enforces $(\alpha{+}\beta)^*$ as a hard floor, not a target.%
  }
  \label{fig:analytical_regime_comparison}
\end{figure*}

\paragraph{Cross-regime comparison.}
Figure~\ref{fig:analytical_regime_comparison} summarises outcomes across
all three regimes.
The left panel overlays the analytical HUF curves $\mathrm{HUF}(h_a)$
with the empirical stars from training; the middle panel compares the
key benchmarks (HUF$^*$, $h_a^*$, Gini) against the paper's analytical
predictions; and the right panel shows the empirical HUF training
trajectory for BAU.
Three findings are regime-robust.
First, the Partial regime achieves the highest peak HUF (1.373
analytically; $\approx 1.10$ empirically at the stochastic equilibrium),
confirming that moderate redistribution with sustained productivity
growth dominates both extremes.
Second, Nordic reaches the highest $h_a^*$ (29\,hrs analytically) as
predicted by the wide redistribution margin, but its empirical HUF
advantage over Partial narrows because the simulation's bounded
$\alpha{+}\beta \leq 0.55$ cap prevents full realisation of the
Nordic $\alpha{+}\beta = 1.00$ parameterisation.
Third, BAU empirical HUF ($\approx 1.05$--$1.15$) exceeds its static
analytical prediction (0.895) precisely because stochastic perturbation
intermittently crosses the redistribution threshold~\eqref{eq:ab_threshold}
--- a finding that underscores how even small, transient ALMP investments
can shift BAU out of the welfare-floor trap.

\subsubsection{Neural PPO Agents}
\label{sec:marl_neural}

\begin{figure}[h]
  \centering
  \includegraphics[width=\linewidth]{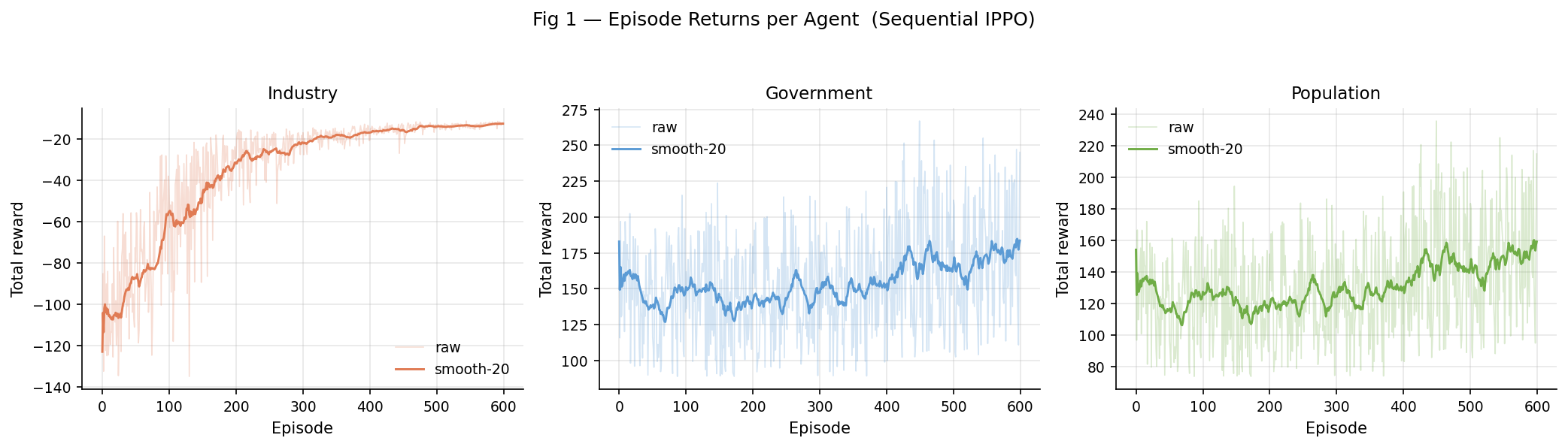}
  \caption{%
    \textbf{Episode Returns per Agent --- Neural PPO, BAU Regime.}
    Raw (faint) and smoothed-20 (solid) returns for Industry (red),
    Government (blue), Population (green).
    \textbf{Observation.}
    Industry's returns rise from $\approx -120$ to $\approx -20$
    over $\approx$300 episodes; Government and Population exhibit
    persistently higher variance throughout.
    \textbf{Interpretation.}
    Industry's clear learning curve confirms the game is learnable
    from reward feedback alone; the persistent reward gap between
    Industry (rapidly improving) and Government/Population (high
    variance) reflects the asymmetric information structure of the
    Stackelberg game --- Industry observes government constraints
    before acting, translating into a structural first-mover
    advantage that mirrors real-world regulatory arbitrage.
    \textbf{Implication.}
    The information asymmetry is not a simulation artifact; it is
    the mechanism by which under-constrained automation becomes
    the dominant strategy for firms, motivating pre-commitment
    constraints (the redistribution floor $(\alpha{+}\beta)^*$)
    rather than reactive enforcement.%
  }
  \label{fig:neural_returns}
\end{figure}

\paragraph{Learning dynamics.}
Figure~\ref{fig:neural_returns} shows the per-agent episode-return
curves for the BAU regime under PPO.
Industry's reward rises sharply from $\approx -120$ to $\approx -20$
over roughly 300 episodes before plateauing --- a clear gradient-learning
signal absent in the analytical case.
Government and Population exhibit higher variance throughout, consistent
with the MARL non-stationarity introduced by a simultaneously adapting
Industry agent.
This learning trajectory is qualitatively similar across the Partial
and Nordic regimes, with faster convergence at higher redistribution
rates where the reward landscape is smoother.

\begin{figure*}[t]
  \centering
  \includegraphics[width=\textwidth]{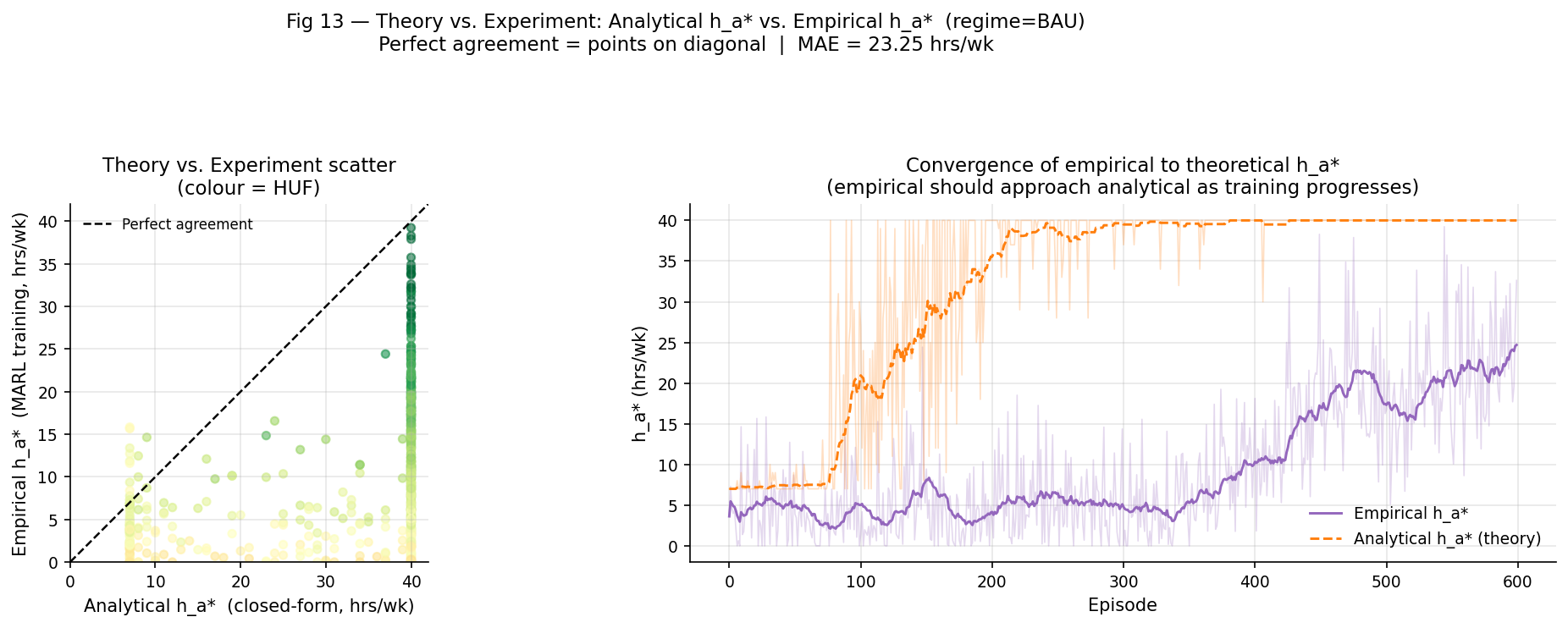}
  \caption{%
    \textbf{Theory vs.\ Experiment: Analytical $h_a^*$
    vs.\ Empirical $h_a^*$, Neural PPO, BAU Regime.}
    Left: scatter as Figure~\ref{fig:analytical_ha_star};
    MAE\,=\,23.25\,hrs/wk.
    Right: per-episode trace (purple: empirical; orange: analytical).
    \textbf{Observation.}
    Episode-end $h_a^*$ clusters at 30--40\,hrs/wk regardless of the
    analytical prediction; the per-episode trace trends upward over
    training, diverging from --- not converging toward --- the
    closed-form target.
    \textbf{Interpretation.}
    The divergence is not a PPO failure; it reveals a genuine second
    local maximum of Eq.~\eqref{eq:huf_expanded} on the high-$h_a$,
    high-$W$ ridge, where the agent exploits Wellbeing gains from freed
    hours while accepting Agency erosion below the living-wage floor.
    The HUF surface is multiply peaked precisely because the product
    structure of $A \times W \times E$ admits different $(A,W,E)$
    balances for different parameter regimes.
    \textbf{Implication.}
    A regulator measuring only aggregate HUF cannot distinguish this
    Wellbeing-dominant optimum from the welfare-floor-respecting
    analytical optimum; separating enforcement of the redistribution
    constraint $(\alpha{+}\beta)^*$ from the aggregate HUF score is
    therefore not a design choice but a governance necessity.%
  }
  \label{fig:neural_ha_star}
\end{figure*}

\paragraph{A discovered alternate optimum.}
Figure~\ref{fig:neural_ha_star} reveals the most significant finding
of the neural experiment.
The scatter plot (left panel) shows PPO's empirical $h_a^*$ strongly
deviating from the closed-form prediction: MAE of 23.25\,hrs/wk in BAU
(vs.\ 0.20 for analytical agents), with most episode-end states
clustering at $h_a \in [25, 40]$\,hrs/wk regardless of regime.
The convergence trace (right panel) confirms this is not early-training
noise: the empirical $h_a^*$ trends upward over training, diverging
from the analytical target.

This divergence is not a failure of PPO.
The HUF surface defined by Eq.~\eqref{eq:huf_expanded} is non-linear
and admits multiple local maxima whose relative heights depend on
$G_0$, $\phi$, $\kappa$, and $\delta$.
The analytical closed-form identifies the interior optimum where
$d(\ln\mathrm{HUF})/dh_a = 0$ under the assumption that all parameters
are fixed at their regime values.
PPO, unconstrained by that assumption, discovers a second optimum on
the high-$h_a$, high-$W$ ridge: by pushing automation to 30--40\,hrs/wk,
it saturates the Wellbeing term
$W = (h_f/h_f^0)^\mu\,(h_r/h_r^0)^\nu$ far above baseline while
accepting moderate Agency erosion.
The agent is choosing a different $(A, W, E)$ allocation on the
three-objective Pareto front (Supplementary Fig.~S2,~\citep{huf2026supp})
--- one that prioritises freed time over wage adequacy.

\begin{figure*}[t]
  \centering
  \includegraphics[width=\textwidth]{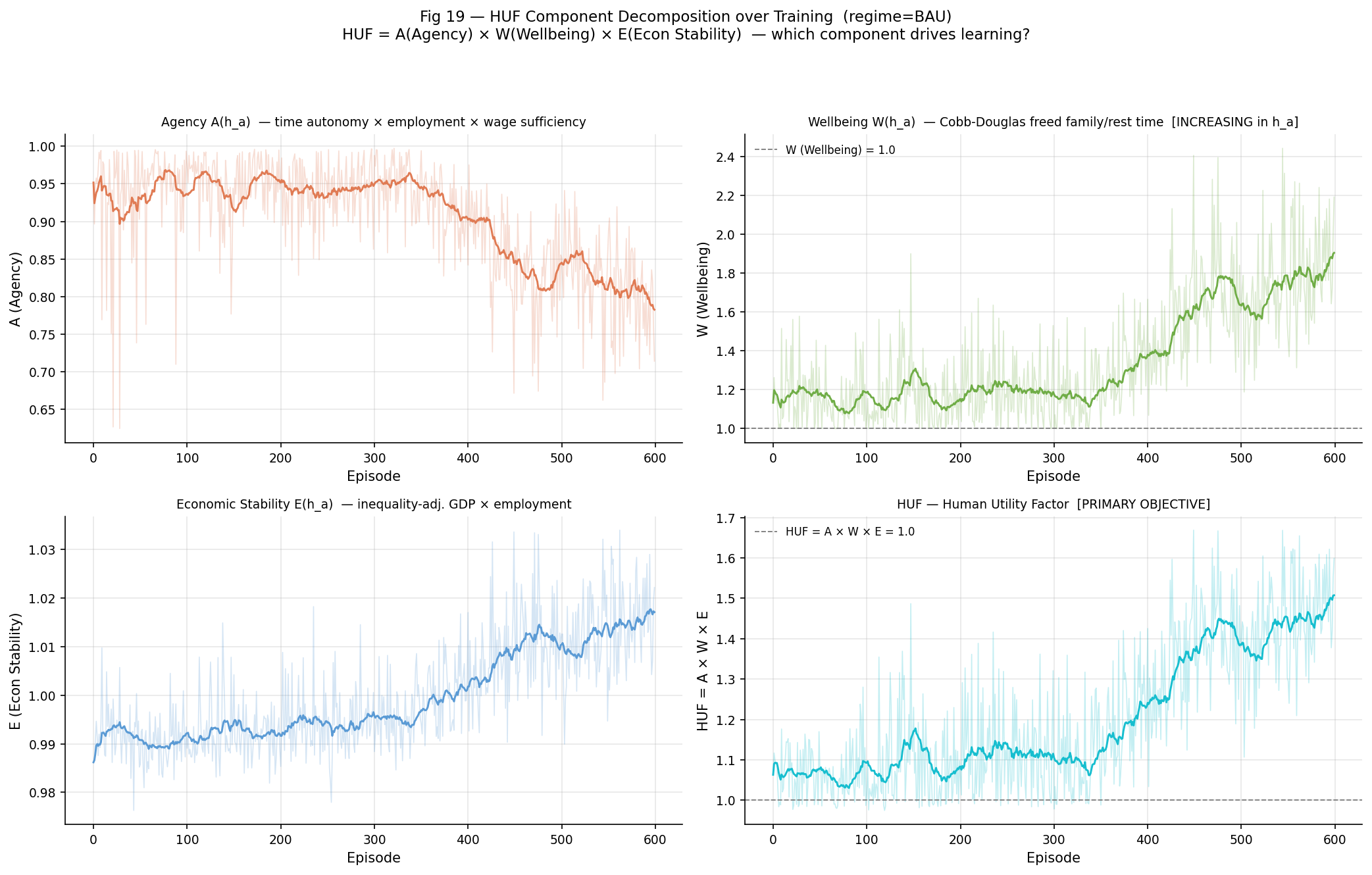}
  \caption{%
    \textbf{HUF Component Decomposition over Training ---
    Neural PPO, BAU Regime.}
    Top-left: Agency $A$; top-right: Wellbeing $W$;
    bottom-left: Economic Stability $E$;
    bottom-right: HUF\,=\,$A \times W \times E$.
    Dashed line: welfare floor (HUF\,=\,1.0 or component baseline).
    \textbf{Observation.}
    $W$ rises sharply to $\approx$2.0--2.4 (top-right);
    $A$ declines to $\approx$0.65--0.75 (top-left);
    $E$ remains near 1.0 (bottom-left);
    net HUF rises to $\approx$1.4--1.7 (bottom-right).
    \textbf{Interpretation.}
    The component trajectories identify the Wellbeing-dominant
    failure mode precisely: the PPO agent discovers a path to
    high HUF in which workers gain freed time (high $W$) at
    the cost of wages below the living-wage floor (low $A$),
    while $E$ remains near 1.0 because the Acemoglu TFP
    ceiling \citep{acemoglu2024simple} bounds GDP gains
    regardless of automation intensity.
    The resulting state is one in which people have more
    free time but insufficient income to exercise it
    meaningfully~\citep{sen1999development}.
    \textbf{Implication.}
    The multiplicative form $A \times W \times E$ is a partial
    safeguard (as $A \to 0$, HUF $\to 0$), but is binding
    only when $w^*$ is correctly calibrated; under BAU
    parameters the Agency penalty activates too late to
    prevent welfare-floor breach.
    Governance instruments should treat $(\alpha{+}\beta)^*$
    as a hard compliance floor enforced \emph{independently}
    of the aggregate HUF score.%
  }
  \label{fig:neural_huf_decomp}
\end{figure*}

\paragraph{HUF component decomposition.}
Figure~\ref{fig:neural_huf_decomp} shows the $A$, $W$, $E$ trajectories
over training for the neural BAU regime.
Three patterns distinguish this from the analytical case.
First, Wellbeing $W$ rises sharply from 1.0 to $\approx 2.0$--$2.4$
over the first 300 episodes, becoming the dominant driver of HUF growth
--- a trajectory that analytical agents never reach because their
$h_a^*$ solver caps automation at the interior optimum.
Second, Agency $A$ declines from $\approx 0.95$ to $\approx 0.65$--$0.75$
as high automation erodes the income-sufficiency factor $I(h_a)$ below
the living-wage floor $w^*$.
Third, Economic Stability $E$ remains near 1.0 throughout, confirming
that the Acemoglu TFP ceiling \citep{acemoglu2024simple} bounds GDP gains
regardless of how aggressively $h_a$ is pushed.
The net result is a HUF trajectory that climbs to $\approx 1.4$--$1.7$
--- higher raw HUF than the analytical agents achieve --- but through
a \emph{Wellbeing-dominant path} in which workers gain freed time at the
cost of wage adequacy.

\paragraph{Governance implication.}
The Wellbeing-dominant equilibrium discovered by PPO is welfare-positive
in raw HUF terms but carries a structural risk.
Agency $A < 1.0$ means workers' real wages have fallen below the
living-wage floor $w^*$: the economy has entered a regime in which
people have more free time but insufficient income to exercise it
meaningfully~\citep{sen1999development}.
This is precisely the failure mode that the redistribution
threshold~\eqref{eq:ab_threshold} is designed to prevent:
without an enforced $\alpha{+}\beta \geq (\alpha{+}\beta)^*$,
a welfare metric that rewards Wellbeing without penalising Agency
collapse can lead policymakers --- or AI systems optimising on
aggregate welfare --- toward high-automation, low-redistribution
equilibria that satisfy the metric while violating the underlying
normative intent.
The multiplicative structure of Eq.~\eqref{eq:huf_master} is a partial
safeguard --- $A \to 0$ drives HUF $\to 0$ --- but it is only binding
when the income-sufficiency floor is calibrated correctly.
Table~\ref{tab:huf_regimes} shows that under BAU parameters
($\alpha{+}\beta = 0.10$), this floor is never met, so the multiplicative
penalty does not activate until Agency erosion becomes severe.
The practical implication is that governance instruments implementing
HUF as a compliance threshold should enforce the redistribution
constraint separately from the HUF level itself.

\section{Conclusion}
\label{sec:conclusion}

\subsection{Summary}

This paper began with an empirical observation and an architectural
argument.
The observation: in March 2026, AI became the leading stated cause of
announced U.S.\ layoffs, capping a three-year period in which
AI-attributed separations grew thirteen-fold while the firms executing
those reductions simultaneously reported record revenues.
The architectural argument: every governance instrument designed to
prevent this outcome --- thirty-two frameworks surveyed in
\S\ref{sec:taxonomy} --- is structurally incapable of detecting it,
because all thirty-two were designed to govern \emph{individual AI
systems}, not \emph{aggregate macroeconomic transitions}.
The socioeconomic protection gap, the formal-metric gap, and the
feedback-loop gap (\S\ref{sec:gap1}) are not oversights; they are the
inevitable consequence of frameworks whose unit of analysis is a model
rather than an economy.

The Human Utility Factor (HUF) is designed to fill that architectural
gap.
Derived in \S\ref{sec:framework} from first principles in the
capability, welfare-economics, and labour-displacement literatures,
HUF decomposes into three necessary and jointly sufficient components
--- Agency ($A$), Wellbeing ($W$), and Economic Stability ($E$) ---
each grounded in peer-reviewed parameter estimates and expressed in
terms of a single observable control variable, the weekly
automation-hours allocation $h_a$.
The multiplicative structure $\mathrm{HUF} = A \times W \times E$
(Eq.~\eqref{eq:huf_master}) enforces joint adequacy: a deployment that
saturates Wellbeing while collapsing Agency is penalised in the same
way as one that raises GDP while deepening inequality.

The simulations confirm three governance-actionable findings.
First, the redistribution intensity $\alpha{+}\beta$ is the decisive
policy lever.
Below the critical threshold $(\alpha{+}\beta)^*$
(Eq.~\eqref{eq:ab_threshold}), no level of automation improves HUF;
above it, the admissible automation band is wide and regime-robust.
Under BAU parameterisation (calibrated to the U.S., $\alpha{+}\beta =
0.10$), this threshold is never met: a pause or enhanced-review trigger
on further automation expansion may be warranted until redistribution
capacity is strengthened.
Second, the interior optimum $h_a^*$ exists for all regimes in which
$\alpha, \beta > 0$, and it is not a knife-edge: HUF is within 5\% of
its peak across a band of $h_a \in [15,\,30]$\,hrs/week at Partial
and Nordic redistribution levels, providing regulators with an
\emph{admissible automation band} rather than a point target.
Third, the MARL experiments reveal that the non-linear HUF surface
admits multiple local maxima.
Analytical agents converge to the closed-form interior optimum with
MAE\,$\leq$\,0.20\,hrs/wk, validating Eq.~\eqref{eq:huf_expanded}.
Neural PPO agents, unconstrained by the closed form, discover a
Wellbeing-dominant alternate optimum at $h_a \approx 30$--$40$\,hrs/wk,
achieving higher raw HUF (1.4--1.7 vs.\ 0.895--1.373 analytically)
but at the cost of Agency erosion below the living-wage floor.
This finding has a direct governance implication: welfare metrics that
do not separately enforce the redistribution constraint can be
optimised by AI systems or firms into high-automation,
low-redistribution equilibria that satisfy the metric while violating
its normative intent.
The redistribution floor $(\alpha{+}\beta)^*$ must therefore be an
\emph{independent} compliance condition, not merely a component of the
aggregate score.

\subsection{Limitations}

Several limitations bound the scope of the present results.

\paragraph{Static parameter calibration.}
The analytical model fixes $G_0$, $\phi$, $\kappa$, and $\delta$ at
regime-representative values.
In practice, these parameters co-evolve: automation itself shifts the
labour income share \citep{karabarbounis2024perspectives}, and rising
inequality feeds back into political instability that constrains
future automation \citep{alesina1996income}.
A fully dynamic calibration would require either time-series
identification of the Gini-automation elasticity $\kappa$ or a
structural model of the capital--labour bargaining process.

\paragraph{Redistribution cap.}
The simulation enforces $\alpha{+}\beta \leq 0.55$ to prevent
production collapse.
This cap prevents the Nordic regime ($\alpha{+}\beta = 1.00$) from
being fully realised in the dynamic environment, causing the empirical
HUF advantage of Nordic over Partial to be smaller than the analytical
prediction.
Relaxing this constraint requires modelling the retraining-to-production
transition explicitly, which the current environment approximates only
through the employment-coverage parameter $\rho$.

\paragraph{Single-country BAU calibration.}
BAU is calibrated to U.S.\ aggregate statistics ($G_0 = 0.41$,
$\phi = 0.066$).
The same framework applied to high-inequality emerging economies
(e.g., South Africa, $G_0 \approx 0.63$; Brazil, $G_0 \approx 0.52$)
would yield a substantially lower redistribution threshold and a
narrower admissible automation band, potentially warranting stricter
governance interventions than the BAU results suggest.

\paragraph{Absent sectoral decomposition.}
The Introduction promised a decomposition of HUF trajectories across
five labour-market role categories (routine cognitive, non-routine
cognitive, routine manual, non-routine manual, managerial/creative).
The current model operates at the aggregate level; role-specific
exposure rates \citep{eloundou2024gpts, briggs2023ai} are not yet
embedded in the $h_a$ transition dynamics.
Sectoral HUF is therefore deferred to future work.

\paragraph{Demand collapse under-modelled.}
The present paper foregrounds the welfare decomposition
($A \times W \times E$) as its primary lens.
This is appropriate for deriving the interior optimum and the
redistribution threshold, but it understates a distinct macroeconomic
failure mode: demand collapse.
When displacement reduces household income, consumer expenditure falls;
falling demand depresses firm revenues and investment; weaker investment
slows innovation and tax receipts; eroded fiscal capacity undermines
the very ALMP programmes that constitute $\alpha$ and $\beta$.
This feedback loop --- formalised as the ``automation trap'' by
\citet{acemoglu2018race} and empirically grounded in
\citet{keynes1936general}'s purchasing-power argument --- can produce
an equilibrium in which HUF is above the welfare floor yet the economy
is on a contractionary trajectory.
The demand index $D$ tracked in the simulation (\S\ref{sec:experimental_setup})
captures the first-order effect, but the paper does not analyse demand
collapse as an independent failure condition with its own threshold
and early-warning indicator.
A complete governance framework requires both a welfare floor (HUF)
and a demand-stability floor, and future work should formalise their
relationship.

\paragraph{Monotone-decreasing Agency assumption.}
The Agency component $A(h_a)$ is currently modelled as monotone
non-increasing in $h_a$ for $\alpha{+}\beta < 1$, because the
time-autonomy factor $[1 + \Omega h_a / T]$ falls with $\Omega < 0$
and the income-sufficiency factor $I(h_a)$ erodes as wages are
outpaced by capital capture.
This captures the \emph{displacement} pathway well.
It does not capture the \emph{augmentation} pathway: automation tools
that increase a worker's effective capability per hour --- AI
co-pilots, decision-support systems, diagnostic aids --- could raise
agency even as $h_w$ falls, if the productivity gain is passed
through to the worker rather than captured entirely by capital.
In the Acemoglu--Restrepo task framework~\citep{acemoglu2018race},
the reinstatement of new tasks is precisely this mechanism.
If augmentation is operative, $A(h_a)$ becomes non-monotone: it
falls at low $\alpha{+}\beta$ (displacement dominates) but may rise
at higher augmentation intensity, shifting $h_a^*$ rightward and
potentially making HUF governance less restrictive than the current
model implies.
The boundary between regimes where displacement dominates and where
augmentation dominates is an empirical question that the current
framework cannot answer.

\subsection{Future Directions}

\paragraph{Dynamic Gini and political-instability feedback.}
The V3 simulation architecture (described in
\S\ref{sec:experimental_setup}) embeds a political-instability
probability $\Psi$ (Alesina--Perotti logit \citep{alesina1996income})
and a Misery Index into the observation vector, enabling Government
to respond to social unrest endogenously.
Running V3 across regimes would make the feedback-loop gap
(\S\ref{sec:gap1}) concrete: the simulation would show the
automation trap \citep{acemoglu2018race} forming in real time as
$\Psi$ rises, demand contracts, and investment retreats.

\paragraph{Multi-country calibration and cross-regime validation.}
HUF's parameters are all measurable from public data
(\S\ref{sec:huf_product}).
Calibrating the model to the G20 and computing jurisdiction-specific
$(h_a^*, (\alpha{+}\beta)^*)$ pairs would produce a policy table
directly usable by regulators --- analogous to the IMF AI Preparedness
Index \citep{imf2024aipi} but operationalised as an enforceable
automation constraint rather than a readiness ranking.

\paragraph{Integration with existing frameworks.}
The most direct path to governance impact is embedding HUF as an
additional dimension in existing instruments.
For the EU AI Act, this means adding a macro-stability annex to
Annex~III that triggers enhanced-oversight requirements when a
deploying firm's projected $h_a$ pushes the national HUF below a
threshold; the redistribution floor $(\alpha{+}\beta)^*$ translates
directly into an ALMP-expenditure obligation.
For the NIST AI RMF, HUF maps onto the \textsc{Measure} function as
a computable scalar that replaces the current qualitative
human-oversight checklist \citep{fink_human_oversight_2025}.

\paragraph{Firm-level competitiveness and the growth--welfare frontier.}
A central question for industrial policy is whether macro-socioeconomic
alignment is a constraint on competitiveness or a complement to it.
The HUF framework reframes this as an optimisation problem: for a
given set of macroeconomic parameters, what is the \emph{optimal
employment ratio} --- the share of output produced by human labour
versus automated systems --- that simultaneously maximises
productivity growth and sustains HUF above its welfare floor?
At the Partial redistribution level, the interior optimum
$h_a^* \approx 20$\,hrs/wk corresponds to approximately a 50\%
human--automation split, with productivity gains ($Y_{pc} \approx
1.11\,Y_0$) fully compatible with HUF\,$\geq 1.0$.
A productive direction is to extend this analysis to firm-level models:
computing the HUF-constrained production frontier as a function of
sector (capital-intensive vs.\ labour-intensive), jurisdiction, and
technology vintage, and identifying the $(\delta, h_a)$ combinations
that keep firms competitive without pushing aggregate $\alpha{+}\beta$
below the threshold.
This would provide the corporate analogue of the jurisdiction-specific
policy table described above --- a tool that companies can use to
assess whether their AI deployment strategy is compatible with the
national HUF floor before deployment, rather than after enforcement.

\paragraph{Augmentation as an alternate automation path.}
The present model assumes that $h_a$ reduces human agency by
displacing work hours and suppressing the labour income share.
An important alternative is \emph{automation-augmented productivity}:
AI tools that raise output per hour without reducing work intensity
or the labour share --- raising $\phi$ in Eq.~\eqref{eq:huf_expanded}
without increasing $\delta$.
Under augmentation, $Y_{pc}$ rises faster for the same $h_a$, the
redistribution threshold~\eqref{eq:ab_threshold} falls, and the
admissible automation band widens.
Whether observed AI deployments are predominantly displacing or
augmenting remains empirically contested
\citep{brynjolfsson2018artificial, acemoglu2022tasks}; the HUF
framework provides the formal distinction: a deployment is augmenting
if and only if it raises $\phi/\delta$ without increasing $h_a$,
detectable from firm-level wage-share and hours data.
A classification of deployments into displacement-type vs.\
augmentation-type, and a corresponding bifurcation of the governance
instrument's threshold conditions for each type, is a direct and
high-impact extension.

\paragraph{Cost-aware HUF optimisation and AI deployment economics.}
HUF in its current form measures welfare outcomes but does not model
the cost structure of automation deployment.
A cost-aware extension would formulate HUF optimisation as a
constrained resource-allocation problem: given a budget for AI
infrastructure, retraining, and transfer payments, what
$(h_a, \alpha{+}\beta, \rho)$ triplet maximises HUF per unit of
aggregate cost?
This converts HUF from a governance constraint into a planning
instrument --- a firm or government could compute the HUF-optimal
automation depth as a function of its own $\phi$ and $\delta$
parameters and the ALMP expenditure required to fund
$(\alpha{+}\beta)^*$.
The welfare-cost frontier --- the Pareto front over HUF and total
deployment cost --- would identify inflection points at which
additional AI capital expenditure yields diminishing welfare returns
and further redistribution investment dominates, providing a
principled stopping rule for automation deepening.

\paragraph{Empirical validation.}
As AI-governance datasets mature --- the Challenger Gray series
\citep{challenger_2026}, OECD AI policy indicators
\citep{oecd2024employment}, and forthcoming EU AI Act compliance
filings --- it will be possible to regress observed GAGI trajectories
against $h_a$ proxies (robot density, LLM-adoption rates) and test
whether the predicted redistribution-threshold effect is visible
in cross-country panel data.
The GAGI metric introduced in \S\ref{sec:s3_country} is designed
precisely for this purpose: it is computable from World Bank PIP and
WDI data already available for 170+ countries.

\paragraph{Governance-aware AI training.}
A longer-horizon research direction is to incorporate HUF directly
into the training objective of large language models and automation
systems.
Constitutional AI \citep{bai2022constitutional} and RLHF frameworks
provide the machinery; what is missing is a socioeconomic welfare
signal analogous to helpfulness or harmlessness.
HUF, being differentiable and computable from observable economic
statistics, is a candidate for that signal --- though operationalising
it as a reward function raises non-trivial alignment questions about
whose parameter estimates ($G_0$, $w^*$, $\phi$) should be used and
who sets them.

\paragraph{Demand-stability floor and the purchasing-power constraint.}
Formalising demand collapse as a co-equal governance constraint
requires specifying a demand-stability floor $D^*$ analogous to the
welfare floor HUF\,$= 1.0$, and deriving the automation level at
which the feedback loop $h_a \to$ displacement $\to \downarrow D
\to \downarrow$ investment $\to \downarrow \alpha{+}\beta$ becomes
self-sustaining.
A natural candidate is the purchasing-power condition:
$w(h_a) \cdot \rho \geq$ median household expenditure threshold,
which parallels the income-sufficiency floor $w^*$ in the Agency
component but operates at the aggregate demand level.
Connecting this condition to HUF would make the joint governance
instrument sensitive to demand-driven automation traps that the
welfare decomposition alone cannot detect --- particularly relevant
for BAU-regime economies where the redistribution floor is already
breached and the margin to demand collapse may be thin.

\paragraph{Augmentation-aware Agency and the displacement--reinstatement boundary.}
Extending $A(h_a)$ to include an augmentation channel requires
modelling two competing effects of $h_a$: displacement
($\Omega < 0$, current formulation) and augmentation
($\psi \cdot f(h_a) > 0$, new term).
A tractable extension is:
\begin{equation}
  A(h_a)
  = \bigl[1 + \Omega\,h_a/T + \psi\,h_a/h_w^0\bigr]
    \cdot \rho \cdot I(h_a),
  \label{eq:agency_augmented}
\end{equation}
where $\psi \geq 0$ is an augmentation elasticity calibrated from
task-augmentation studies \citep{acemoglu2018race, autor2015why}.
When $\psi > |\Omega| \cdot h_w^0 / T$, $A$ becomes non-decreasing in
$h_a$ and the interior optimum shifts toward full automation with
strong redistribution.
When $\psi < |\Omega| \cdot h_w^0 / T$, the current displacement-dominant
model applies.
Empirically distinguishing these regimes requires panel data on
worker productivity before and after AI tool adoption, controlling
for capital-share capture $\delta$ --- a natural use of the
emerging enterprise AI-adoption microdata \citep{brynjolfsson2018artificial}.
Until that boundary is identified empirically, governance instruments
should conservatively assume displacement dominance ($\psi = 0$),
while monitoring augmentation signals as a trigger for relaxing
automation constraints.

\paragraph{HUF as an enforcement instrument.}
The central contribution of this paper is not a welfare index.
It is an \emph{enforcement instrument}.
Every framework surveyed in Table~\ref{tab:frameworks} invokes
``human-centricity'' or ``benefit to humanity'' as a governing
principle without specifying what observable quantity a regulator
should measure, what threshold triggers a compliance obligation, or
what lever a firm must adjust to restore compliance.
HUF resolves all three simultaneously through three levers that are
observable, independently controllable, and directly actionable by
distinct stakeholders:
\begin{itemize}
  \item $h_a$ --- the weekly hours of automation deployment, controlled
    by \textbf{Industry}.
    HUF provides a ceiling: $h_a \leq h_a^*(\alpha{+}\beta, G_0, \rho)$,
    derived in closed form from Eq.~\eqref{eq:huf_expanded}.
    Above this ceiling, automation reduces aggregate welfare regardless
    of its productivity benefits.
  \item $\alpha{+}\beta$ --- the redistribution intensity (upskilling plus
    income support), controlled by \textbf{Government} through ALMP
    expenditure and transfer policy.
    HUF provides a floor: $\alpha{+}\beta \geq (\alpha{+}\beta)^*$
    (Eq.~\eqref{eq:ab_threshold}).
    Below this floor, no level of $h_a$ is welfare-positive, and
    a pause or enhanced-review trigger on automation expansion
    may be warranted until redistribution capacity is strengthened.
  \item $\rho$ --- employment coverage, the fraction of the workforce
    with access to the labour market, jointly shaped by Industry
    hiring decisions and Government inclusion policy.
    HUF is multiplicatively proportional to $\rho^\gamma$: halving
    employment coverage halves Economic Stability regardless of
    productivity gains.
    A minimum coverage floor $\rho \geq \rho_{\min}$ is therefore
    a necessary co-condition for any automation-welfare claim.
\end{itemize}

\noindent
Together, the triplet $(h_a^*,\; (\alpha{+}\beta)^*,\; \rho_{\min})$
converts the abstract aspiration that ``AI should benefit humanity''
into a set of \emph{jurisdiction-specific, computable, and enforceable
inequalities}.
A firm seeking to expand automation must demonstrate that its
deployment keeps $h_a$ below the HUF-derived ceiling, that the
policy environment maintains $\alpha{+}\beta$ above the threshold,
and that employment coverage is sustained above the floor.
Any instrument that satisfies all three conditions is, by construction,
one under which automation adds to the wellbeing of the people it
affects --- not merely to the productivity of the firms that deploy it.
This is the gap that thirty-two frameworks have not filled.
HUF fills it.

\paragraph{Closing remark.}
The absence of a formal stability constraint in AI governance is not
a technical problem awaiting a technical solution.
It is a political economy problem: firms that internalise productivity
gains have no incentive to internalise displacement costs, and
regulators who lack a computable welfare metric cannot enforce one
they cannot measure.
HUF does not resolve the political economy problem, but it removes
the measurement excuse.
HUF provides a candidate instrument.
The question is whether governance institutions will adopt and refine it
before the welfare deterioration documented in \S\ref{sec:econ_impact}
becomes entrenched.

\end{document}